\documentclass[usenatbib]{mnras}

\usepackage{newtxtext,newtxmath}

\usepackage[T1]{fontenc}
\usepackage{ae,aecompl}
\usepackage{times}

\usepackage{graphicx}	

\usepackage{makecell}

\def \MSUN{{\rm M}_{\odot}}

\title[SHMR of TNG satellites and centrals]{The distinct stellar-to-halo mass relations of satellite and central galaxies: insights from the IllustrisTNG simulations}

\author[C. Engler et al.]{Christoph Engler$^{1,2}$\thanks{E-mail: engler$@$mpia-hd.mpg.de},
Annalisa Pillepich$^{2}$, Gandhali D. Joshi$^{2}$, Dylan Nelson$^{3}$,
\newauthor
Anna Pasquali$^{1}$, Eva K. Grebel$^{1}$, Thorsten Lisker$^{1}$, Elad Zinger$^{2}$, 
Martina Donnari$^{2}$, 
\newauthor
Federico Marinacci$^{4}$, Mark Vogelsberger$^{5}$, and Lars Hernquist$^{6}$
\\ \\
$^{1}$Astronomisches Rechen-Institut, Zentrum f\"{u}r Astronomie der Universit\"{a}t Heidelberg, M\"{o}nchhofstra\ss e 12-14, 69120 Heidelberg, Germany\\
$^{2}$Max-Planck-Institut f\"{u}r Astronomie, K\"{o}nigstuhl 17, 69117 Heidelberg, Germany\\
$^{3}$Max-Planck-Institut f\"{u}r Astrophysik, Karl-Schwarzschild-Str. 1, 85741 Garching, Germany\\
$^{4}$Department of Physics \& Astronomy, University of Bologna, via Gobetti 93/2, 40129 Bologna, Italy\\
$^{5}$Kavli Institute for Astrophysics and Space Research, Massachusetts Institute of Technology, Cambridge, MA 02139, USA\\
$^{6}$Harvard-Smithsonian Center for Astrophysics, 60 Garden Street, Cambridge, MA 02138, USA
}

\date{Accepted 2020 November 6. Received 2020 November 6; in original form 2020 February 25}

\pubyear{2020}

\begin{document}
\label{firstpage}
\pagerange{\pageref{firstpage}--\pageref{lastpage}}
\maketitle

\begin{abstract}
We study the stellar-to-halo mass relation (SHMR) for central and satellite galaxies with total dynamical masses above $10^{10.5}~\rmn{M}_\odot$ using the suite of cosmological magneto-hydrodynamical simulations IllustrisTNG. In particular, we quantify environmental effects on satellite populations from TNG50, TNG100, and TNG300 located within the virial radius of group- and cluster-like hosts with total masses of $10^{12-15.2}~\rmn{M}_\odot$. At fixed stellar mass, the satellite SHMR exhibits a distinct shift towards lower dynamical mass compared to the SHMR of centrals. Conversely, at fixed dynamical mass, satellite galaxies appear to have larger stellar-to-total mass fractions than centrals by up to a factor of a few. The systematic deviation from the central SHMR is larger for satellites in more massive hosts, at smaller cluster-centric distances, with earlier infall times, and that inhabit higher local density environments; moreover, it is in place already at early times ($z\lesssim2)$. Systematic environmental effects might contribute to the perceived galaxy-to-galaxy variation in the measured SHMR when galaxies cannot be separated into satellites and centrals. The SHMR of satellites exhibits a larger scatter than centrals (by up to $\sim 0.8$ dex), over the whole range of dynamical mass. The shift of the satellite SHMR results mostly from tidal stripping of their dark matter, which affects satellites in an outside-in fashion: the departure of the satellite SHMR from the centrals' relation diminishes for measurements of dynamical mass in progressively smaller apertures. Finally, we provide a family of fitting functions for the SHMR predicted by IllustrisTNG.
\end{abstract}




\section{Introduction}
\label{sec:intro}

While the formation and evolution of galaxies is governed by a blend of both nature and nurture, their environment determines which dominates. Whether a galaxy spends its lifetime in the field or whether it is bound to a more massive group or cluster environment sets it on a different evolutionary path. Galaxy clusters -- the most massive, gravitationally collapsed structures in the Universe -- offer both large galaxy populations as well as a range of environmental processes that leave their imprint on infalling satellite galaxies. In group or cluster environments, any galaxy can become subject to galaxy-galaxy interactions such as harassment \citep{moore1996, moore1998} -- high-velocity encounters driving morphological transformation -- or various interactions with the host halo's potential: in a starvation scenario, gas accretion from the surrounding halo into the galaxy is cut off. Star formation continues for an extended period of time until the galaxy's gas reservoirs have been exhausted \citep{larson1980, balogh2000, kawata2008, wetzel2013}. Ram pressure stripping \citep{gunngott1972} deprives galaxies in the intracluster or intragroup medium of their gas, thereby removing the reservoirs for the formation of new stars and rapidly quenching the galaxies \citep[e.g.][]{tonnesen2007, kenji2014, fillingham2016, simpson2018} -- possibly after a final, ram pressure-induced episode of enhanced star formation \citep{vulcani2018, safarzadeh2019}. Interactions between the cold interstellar and the hot intergalactic medium can cause the interstellar medium's temperature to increase rapidly, followed by evaporation and removal of the gas therein \citep[e.g.][]{cowie1977, boselli2006}. Finally, tidal stripping in the host cluster potential can remove the surrounding dark matter haloes of satellite galaxies, stars from their outskirts, produce tidal tails or even lead to their disruption \citep[e.g.][]{merritt1983, barnes1992}.

Due to these processes, galaxy populations in groups and clusters are distinct from their counterparts in the field. Satellite morphologies and star formation activity correlate with the density of their surroundings, resulting in high-density environments containing higher fractions of early-type galaxies \citep{einasto1974, oemler1974, dressler1980, binggeli1987, lisker2007, grebel2011} and enhanced quenched fractions \citep{lewis2002, vanderwel2010, spindler2018}. This is directly observable in a higher red fraction for galaxies in high-density environments \citep{font2008, lisker2008, vandenbosch2008, peng2010, prescott2011}. However, these environmental effects are neither restricted to the central regions of clusters, nor to present-day times or the satellites' present-day environment. Ram pressure stripping can already act on satellites that are several virial radii outside of the host cluster \citep{balogh1999, vonderlinden2010, bahe2013, zinger2018}. Preprocessing in previous, group-like hosts can already result in tidal stripping and significant mass loss of a galaxy's surrounding dark matter halo \citep{joshi2017, han2018}. Even after infall into a cluster, such groups can stay bound and still exert their individual influence on satellites. Although groups usually get dispersed after the first pericentric passage, former member galaxies can still appear related at later times -- either in their general properties or their position in phase space \citep{vijayaraghavan2013, lisker2018}. Apart from sharing their time of infall, such galaxies experience similar degrees of tidal mass loss or exhibit quenching and enrichment to similar extents \citep{smith2015, rhee2017, pasquali2019}.

Cosmological simulations offer a convenient way to study the formation and evolution of galaxies in different environments -- either by using pure dark matter simulations, such as Millennium or Millenium II \citep{springel2005, boylan-kolchin2009}, in combination with semi-analytic models \citep[e.g.][]{guo2011}, or by using cosmological hydrodynamical simulations, such as EAGLE \citep{schaye2015}, Horizon-AGN \citep{dubois2014}, or Illustris \citep{vogelsberger2014a, nelson2015}. These simulations allow for detailed studies of environmental effects on satellite galaxies, comparisons of late- and early-type galaxy populations at different epochs, or the impact of infall time on the enrichment of galaxies and their mass-metallicity relation \citep{weinmann2011, lisker2013, sales2015, engler2018}.

The evolution of galaxies is tightly correlated with the mass of their dark matter halo. Galaxy properties, most fundamentally stellar mass or luminosity, are tightly linked to halo mass and the depth of the halo potential. For central galaxies, this stellar-to-halo mass relation (SHMR) has been well constrained -- either using HI line widths \citep{tully1977}, abundance matching techniques \citep[e.g.][]{nagai2005, behroozi2010, behroozi2013, moster2010, moster2013, allen2019}, weak lensing measurements \citep[e.g.][]{mandelbaum2006, huang2019, sonnenfeld2019}, or simulations \citep[e.g.][]{pillepich2018b, matthee2017}. Other methods of constraining halo properties include X-ray observations \citep[e.g.][]{lin2003, lin2004, yang2007, kravtsov2018}, employing galaxy kinematics, stellar velocities, or planetary nebulae as tracers for the halo potential \citep[e.g.][]{erickson1987, ashman1993, peng2004, vandenbosch2004}, or by measuring the mass or abundance of globular clusters \citep[e.g.][]{spitler2009, forbes2018, prole2019}. For centrals, the SHMR's scatter has been found to correlate with the assembly and the hierarchical growth of massive galaxies, as well as their large-scale environment or halo characteristics, such as its concentration or its growth rate \citep{tonnesen2015, gu2016, goldenmarx2018, goldenmarx2019, feldmann2019, bradshaw2020}.

However, compared to central galaxies, the SHMR of satellites has been found to show significant deviations due to environmental influence \citep{rodriguezPuebla2012, rodriguezPuebla2013, tinker2013, hudson2015, vanuitert2016, bahe2017, sifon2018, buck2019, dvornik2020}.
Here, tidal stripping removes large parts of a satellite's surrounding dark matter. This process already becomes active outside of the host's virial radius \citep{reddick2013, behroozi2014, smith2016} and drives satellite galaxies off of their original position in the SHMR \citep{niemiec2017, niemiec2019}. However, \cite{joshi2019} showed that the dark matter subhaloes of satellites are already subject to tidal stripping as part of preprocessing in groups. During this process, the galaxy itself can still continue its star formation. This suggests that preprocessing plays a significant role in causing the scatter in the SHMR of satellites. But how does the SHMR of satellites vary for different host environments? How do lower-mass groups or massive galaxy clusters influence the SHMR's scatter? And how can we characterise galaxy environment for satellites inside these hosts?

In this study, we examine the SHMR using the cosmological magneto-hydrodynamical simulation suite IllustrisTNG \citep{marinacci2018, naiman2018, nelson2018, pillepich2018b, springel2018, nelson2019a, pillepich2019, nelson2019b}. Here, at least 31 per cent of cluster galaxies with stellar mass above $10^9$ M$_\odot$ have been subject to ram pressure stripping: this is observable in gaseous tails tracing the infalling galaxies and turning them into Jellyfish galaxies (\citealp{yun2019} with TNG, or observationally e.g. \citealp{mcpartland2016, jaffe2018}). While there are still apparent deviations from observations in the star-forming main sequence at earlier times, the amount of quiescent galaxies at intermediate stellar mass are in better agreement with observations than previous models \citep{donnari2019, donnari2020b}. Furthermore, satellite galaxies exhibit enhanced metallicities due to chemical preprocessing (\citealp{gupta2018}, or observationally e.g. \citealp{grebel2003, pasquali2010}).

In this paper we study the SHMR in IllustrisTNG by comparing central and satellite galaxies selected above the same minimum total dynamical mass ($M_\rmn{dyn} \geq 10^{10.5} \text{ M}_\odot$). We focus mostly on $z=0$ but comment on the redshift evolution of the relations and their galaxy-to-galaxy variations up to $z\sim2$. We define a number of environmental parameters and examine their effects on satellite galaxies in groups and clusters, their locus in the SHMR and the scatter in stellar mass. The combination of all the runs of the IllustrisTNG suite allows us to explore an unprecedented dynamical range of satellite and host masses. The nature of the simulations (uniform volumes instead of e.g. zoom-in simulations) allows us to replicate the shape of the mass distributions of host haloes and their satellite galaxies closely, as compared to how they emerge in the real Universe. The paper is structured as follows: in Section~\ref{sec:methods} we describe the IllustrisTNG simulations in detail, define our selection of galaxies, and introduce the parameters we adopt to characterise their environment. We present our results in Section~\ref{sec:res}: the SHMR of centrals and satellites, its scatter as a function of dynamical mass, and the influence of various environmental quantities on the SHMR of satellite galaxies. In Section~\ref{sec:disc}, we discuss the processes that act on satellites after infall into a more massive environment, as well as their transition from the SHMR of centrals. Furthermore, we provide a series of fitting functions for the SHMR in IllustrisTNG and examine the limitations of halo finders and resolution effects, as well as how they affect our results. Finally, we summarise our work in Section~\ref{sec:conc}.

\section{Methods}
\label{sec:methods}

\subsection{IllustrisTNG}
\label{sec:sim}

The results presented in this paper are based on data from IllustrisTNG\footnote{\url{http://www.tng-project.org/}}, \textit{The Next Generation} suite of state-of-the-art magneto-hydrodynamical cosmological simulations of galaxy formation \citep{marinacci2018,naiman2018,nelson2018,pillepich2018b,springel2018}. Building on the success of its predecessor Illustris \citep{vogelsberger2014a,vogelsberger2014b,genel2014,nelson2015,sijacki2015}, IllustrisTNG follows the same fundamental approach but includes improved aspects and novel features in its galaxy formation model and expands its scope to several simulated volumes and improved resolution. The models for galaxy formation include physical processes such as gas heating by a spatially uniform and time-dependent UV background, primordial and metal-line gas cooling, a subgrid model for star formation and the unresolved structure of the interstellar medium \citep{springel2003}, as well as models for the evolution and chemical enrichment of stellar populations, which track nine elements (H, He, C, N, O, Ne, Mg, Si, Fe) in addition to europium and include yields from supernovae Ia, II, and AGB stars \citep{vogelsberger2013,torrey2014}. Furthermore, IllustrisTNG incorporates improved feedback implementations for galactic winds caused by supernovae as well as accretion and feedback from black holes. In particular, depending on accretion, black hole feedback occurs in two modes: low accretion rates result in purely kinetic feedback while high accretion rates invoke thermal feedback \citep{weinberger2017}. Galactic winds are injected isotropically and the wind particles' initial speed scales with the one-dimensional dark matter velocity dispersion \citep{pillepich2018a}. Magnetic fields are amplified self-consistently from a primordial seed field and follow ideal magnetohydrodynamics \citep{pakmor2013}. The TNG simulations were run using the moving mesh code \textsc{Arepo} \citep{springel2010}. Here, concepts from adaptive mesh refinement and smooth particle hydrodynamics are combined to create an unstructured, moving Voronoi tessellation. IllustrisTNG follows the $\Lambda$CDM framework, adopting cosmological parameters according to recent constraints from Planck data: matter density $\Omega_\rmn{m} = 0.3089$, baryonic density $\Omega_\rmn{b} = 0.0486$, cosmological constant $\Omega_\Lambda = 0.6911$, Hubble constant $h = 0.6774$, normalisation $\sigma_8 = 0.8159$, and spectral index $n_\rmn{s} = 0.9667$ \citep{planck2016}.

The TNG suite simulates three different cubic volumes with side lengths of approximately 50 Mpc, 100 Mpc, and 300 Mpc, referred to as TNG50, TNG100, TNG300, respectively. Recently finished, TNG50 offers a higher mass resolution than the other volumes and a detailed look at galaxies and their properties \citep{nelson2019b, pillepich2019}. While TNG300 has a lower resolution, its greater volume provides large statistical samples of galaxies and dense environments, including about 270 galaxy clusters exceeding $10^{14}~\MSUN$ \citep[see e.g.][and see Section~\ref{sec:sample} for the definition of cluster/host mass]{pillepich2018b}. The intermediate volume TNG100 adopts the same initial conditions as the original Illustris simulation and provides both statistical samples of galaxies in field, groups, and clusters, as well as an adequate mass resolution to study these objects. In this paper, we study a combined sample of galaxies from all simulations of the IllustrisTNG suite: TNG300, TNG100, and TNG50. Specifics on each simulation are summarised in Table~\ref{tab:sims}.

\begin{table}
    \centering
    \begin{tabular}{l c c c c}
         \hline \hline
         Simulation & $L_\rmn{box}$ [Mpc] & $N_\rmn{DM}$ & $m_\rmn{DM}$ [M$_\odot$] & $m_\rmn{b}$ [M$_\odot$]\\ \hline
         TNG300 & $302.6$ & $2500^3$ & $5.9 \times 10^7$ & $1.1 \times 10^7$ \\
         TNG100 & $110.7$ & $1820^3$ & $7.5 \times 10^6$ & $1.4 \times 10^6$\\ 
         TNG50 & $51.7$ & $2160^3$ & $4.5 \times 10^5$ & $8.5 \times 10^4$\\ \hline
    \end{tabular}
    \caption{Simulation details for TNG300, TNG100, and TNG50 -- the flagship runs of the IllustrisTNG project used in this work. Parameters include the side length of the simulation box $L_\rmn{box}$, the number of dark matter particles $N_\rmn{DM}$, as well as the mass of both dark matter and baryonic particles $m_\rmn{DM}$ and $m_\rmn{b}$, the latter representing the typical stellar particle mass.}
    \label{tab:sims}
\end{table}

\begin{table}
    \centering
    \begin{tabular}{l r r r}
        \hline \hline
        Host $M_\rmn{200c}$ & TNG300 & TNG100 & TNG50 \\ 
        \hline
        $10^{12} - 10^{13}~\rmn{M}_\odot$ & 35,464 & 1,708 & 183\\
        $10^{13} - 10^{14}~\rmn{M}_\odot$ & 3,453 & 168 & 23\\
        $10^{14} - 10^{14.5}~\rmn{M}_\odot$ & 239 & 11 & 1 \\
        $10^{14.5} - 10^{15.2}~\rmn{M}_\odot$ & 41 & 3 & 0 \\ 
        \hline
    \end{tabular}
    \caption{Number of host haloes in TNG300, TNG100, and TNG50 at $z=0$. We divide haloes into bins of virial mass $M_\rmn{200c}$ for all simulation volumes to account for lower-mass groups and massive galaxy cluster environments.}
    \label{tab:samples_hosts}
\end{table}
\subsection{Galaxy sample and environmental properties}
\label{sec:sample}

We study galaxies between $z=0$ and $z=2$ over a large range of mass, by limiting our sample to objects with a total dynamical mass of $M_\rmn{dyn} \geq 10^{10.5} \text{ M}_\odot$ in order to touch on the dwarf regime without getting into conflict with the simulation's resolution limit. We define dynamical mass as the sum of all gravitationally bound resolution elements identified by the \textsc{subfind} algorithm \citep[][and see Section~\ref{sec:mass} for more details on our fiducial mass measurements]{springel2001, dolag2009}. Within a larger particle group -- haloes determined by a friends-of-friends (FoF) algorithm -- \textsc{subfind} detects substructures of particles as locally overdense regions that are gravitationally self-bound. The \textsc{subfind} catalogue returns central as well as satellite subhaloes. Centrals are gravitationally bound objects whose position coincides with the centre of FoF haloes, i.e. the minimum of the gravitational potential. This includes both brightest cluster galaxies at the high-mass end or field galaxies at lower masses. Any other \textsc{subfind} objects within a FoF halo are called satellites. A priori, satellite subhaloes may be either dark or luminous (i.e. contain a non-vanishing number of stellar particles, in which case they are called satellite galaxies) and can be members of their parent FoF group regardless of their distance from the centre. In this work, we only consider luminous subhaloes (i.e. with at least one stellar particle) and include both centrals and satellites in our sample.

Since we are particularly interested in satellites in groups and clusters, i.e. environments that are expected to leave some sort of imprint on them, we only consider satellite galaxies in hosts of $M_\rmn{host} \geq 10^{12} \text{ M}_\odot$ in the following sections -- with hosts being the FoF halo the respective satellite galaxy inhabits. As host mass $M_\rmn{host}$, we use its virial mass $M_\rmn{200c}$ -- the total mass of a sphere around the FoF halo's centre with a mean density of 200 times the critical density of the universe. Furthermore, we define satellites as only those galaxies found within the virial radius $R_\rmn{200c}$ of their FoF hosts at the time of observation. While this excludes backsplash galaxies -- galaxies which are currently located outside the virial radius or the FoF halo after experiencing a first infall and their first pericentric passage -- we have verified that their inclusion would not alter our results in a significant manner by using the catalogs from \cite{zinger2020}. However, not all satellites represent actual galaxies. Some correspond to fragmentations and clumps within other galaxies due to e.g. disk instabilities that \textsc{subfind} identified as independent objects. Since these non-cosmological objects contain little to no dark matter, we only regard subhaloes with a dark matter mass fraction (to total mass, i.e. including gas too) of at least 10 per cent in order to remove these clumps \citep[see discussion section 5.2 in][]{nelson2019a}. Additionally, we require satellites to reside at a cluster-centric distance of at least $0.05~R_\rmn{200c}$. This way we avoid the innermost host regions, where the identification of subhaloes can become troublesome due to the large density of their surroundings.

At $z=0$, these selection criteria leave us with a sample of 62,253 (3,373; 307) satellite galaxies in TNG300 (TNG100; TNG50). However, groups and clusters can act as very different environments. They cover a large range of mass and act differently on satellite galaxies. In order to compare these effects, we further divide the satellites into subsamples according to the virial mass of their host haloes. We summarise the demographics of available host haloes and the number of galaxies in each subsample for TNG300, TNG100 and TNG50 in Tables~\ref{tab:samples_hosts} and \ref{tab:samples_galaxies}.
\begin{table}
    \centering
    \small
    \begin{tabular}{l r r r}
        \hline \hline
        Sample & TNG300 & TNG100 & TNG50\\ \hline
        Centrals & 624,682 & 41,824 & 4,358\\
        Satellites, $M_\rmn{host} \geq 10^{12}~\rmn{M}_\odot$ & 62,258 & 3,373 & 307\\
        Satellites, $M_\rmn{host} = 10^{12} - 10^{13}~\rmn{M}_\odot$ & 22,347 & 1,121 & 124 \\
        Satellites, $M_\rmn{host} = 10^{13} - 10^{14}~\rmn{M}_\odot$ & 24,662 & 1,367 & 183 \\
        Satellites, $M_\rmn{host} = 10^{14} - 10^{14.5}~\rmn{M}_\odot$ & 9,867 & 556 & 40 \\
        Satellites, $M_\rmn{host} = 10^{14.5} - 10^{15.2}~\rmn{M}_\odot$ & 5,382 & 329 & 0 \\ \hline
    \end{tabular}
    \caption{Galaxy samples in TNG300, TNG100, and TNG50 at $z=0$. This includes centrals (top row) and satellites in group and cluster environments. We study subhaloes with total dynamical masses of $M_\rmn{dyn} \geq 10^{10.5}~\rmn{M}_\odot$. This limit translates in effect into galaxies with stellar mass of about a few $10^8~\rmn{M}_\odot$ and above. Satellites are defined as galaxies within their host's virial radius $R_\rmn{200c}$.}
    \label{tab:samples_galaxies}
\end{table}
Beyond host mass, we use more specific quantities to assess the immediate environment of satellite galaxies. These are:
\begin{enumerate}
\item \textit{Cluster-centric distance:} distance to the central galaxy of the host halo. The gravitational potential and tidal forces grow stronger towards the cluster centre \citep[e.g.][]{gnedin1999}. Cluster-centric distances are given in units of the host's virial radius. 

\item \textit{Infall times:} we use the satellite galaxies' first infall through the virial radius $R_\rmn{200c}$ of their present-day host's main progenitor to account for the duration over which they have been subject to external effects.

\item \textit{Local luminosity density:} local luminosity density describes the satellites' immediate surroundings and their proximity to other galaxies. We generalize the approach in \cite{sybilska2017} for a larger range in host mass: for each satellite, we consider other galaxies within a fixed three-dimensional aperture, sum up their r-band luminosities and divide by the volume of the sphere. As radius for the aperture we use 10 per cent of the host's virial radius. Furthermore, we only take subhaloes with a stellar mass of at least $10^{9}~\rm{M}_\odot$ (within twice the stellar half-mass radius) into account in order to ensure an appropriate level of resolution for neighbouring galaxies.
\end{enumerate}

Furthermore, we discuss an alternative sample of satellites in Sections~\ref{sec:massloss} and \ref{sec:infalltimes} in addition to our fiducial selection. In this case, we do not limit satellite galaxies by their present-day dynamical mass at $z=0$ but by their peak dynamical mass, to all satellites that have ever reached $M_\rmn{dyn} \geq 10^{10.5}~\rmn{M}_\odot$ throughout their lifetime. This enables us to analyse the impact of environment on different mass components of present-day satellite populations over a wider range of masses.

\subsection{Mass measurements}
\label{sec:mass}
 
Throughout this work we compare different operational definitions of a galaxy's stellar mass and total dynamical mass. In either case, we account only for those stellar particles or resolution elements that are labelled as gravitationally bound to a galaxy according to the \textsc{subfind} algorithm. The results presented in this analysis therefore rely on the accuracy of \textsc{subfind} \citep[e.g.][]{ayromlou2019}. Other halo finders might return somewhat different mass measurements and we comment on this in the Discussion Section~\ref{sec:halofinder}. While we do not expect our qualitative findings to change, quantitative results might be subject to biases. Furthermore, we impose additional 3D radial cuts for mass measurements, which can either represent galaxy-specific structural properties or simply correspond to fixed 3D apertures. Notice that for our galaxy sample and analysis we do not employ halo mass descriptors such as $M_\rmn{200c}$ or other spherical-overdensity definitions as these would only be useful for centrals and would not be meaningful for satellites -- since the latter merely represent slight enhancements on the overall background density distributions dominated by their underlying cluster or group hosts. 

Our fiducial choices for galaxy masses read as follows:
\begin{itemize}
    \item $M_*$: a galaxy's stellar mass is the sum of the mass of all the gravitationally bound stellar particles found within twice the stellar half-mass radius $R_{1/2}^*$ from the galaxy centre. While the stellar half-mass radius is calculated from all gravitationally bound stellar particles in the subhalo as identified by \textsc{subfind}, we limit stellar mass in this way since we are specifically interested in the galaxy's main body, not its diffuse outskirts.
    \item $M_\rmn{dyn}$: a galaxy's total dynamical mass is the sum of all gravitationally bound resolution elements (dark matter, stellar and black hole particles, gas cells) as identified by \textsc{subfind}.
\end{itemize}
For other apertures, we follow the approach in \cite{pillepich2018b} and consider total and stellar masses within $100~\rmn{pkpc}$ (physical kpc), $30~\rmn{pkpc}$, $10~\rmn{pkpc}$ and $5~\rmn{pkpc}$. However, we still only consider particles that are gravitationally bound to the subhalo. We choose these apertures to take different galaxies and their components into consideration: depending on the mass of subhaloes, stellar half-mass radii can range from a few kpc in Milky Way-like haloes to tens of kpc for central galaxies of group environments. Furthermore, most of the stellar mass of Milky Way-like galaxies is enclosed within $30~\rmn{kpc}$ -- this aperture provides stellar mass estimates roughly comparable with observational measurements within Petrosian radii \citep{schaye2015}. We include stellar mass measurements in $5~\rmn{pkpc}$ to account for less massive galaxies in our sample. 
Importantly, distinguishing among different mass definitions allows us to characterise how different parts of galaxies are affected by environmental effects such as tidal stripping, and thus how the different mass definitions affect the description and quantification of the stellar-to-halo mass relations for centrals and satellites separately.

Note that unless otherwise stated, we define $M_\rmn{dyn}$ as the dynamical mass at the \textit{present day} since we specifically aim to investigate differences of satellite to central galaxies caused by their environment. Other studies have characterised satellite subhalo mass as peak masses, i.e. before they became subject to environmental effects. In this case, most of the differences we find in this work comparing the SHMRs of centrals and satellites would be mitigated \citep[e.g.][]{shi2020}.

In order to account for discrepancies resulting from resolution effects between the three simulation volumes, we rescale stellar mass in both TNG300 and TNG100 to TNG50. Typically, this results in an increase in stellar mass by a factor of $\sim 2$ ($\sim 1.5$) in TNG300 (TNG100). However, it reaches up to a factor of a few at the low-mass end. These versions are denoted as rTNG300 and rTNG100, respectively. The rescaling process is described in detail in Appendix~\ref{sec:resc}.

\subsection{Functional form and fit of the stellar-to-halo mass relation}
\label{sec:fit}

In what follows, we quantify the relationship between total dynamical mass $M_\rmn{dyn}$ and stellar mass $M_*$ of galaxies by either plotting the latter vs. the former or by plotting the ratio of the stellar to the dynamical mass vs. the dynamical mass. We use the expression stellar-to-halo mass relation (SHMR) for either form and describe the latter by adopting a parametrization similar to \cite{moster2010,moster2013}:
\begin{equation}
    \frac{M_*}{M_\rmn{dyn}} = 2 N\left[\left(\frac{\log M_\rmn{dyn}}{\log M_1}\right)^{-\beta} + \left(\frac{\log M_\rmn{dyn}}{\log M_1}\right)^{\gamma}\right]^{-1},
    \label{eq:mosterfit}
\end{equation}
Here, we modify the original equation from \cite{moster2010, moster2013} to employ logarithmic masses. The four free parameters correspond to the normalisation of the stellar-to-halo mass ratio $N$, a characteristic mass $M_1$, and the two slopes at the low- and high-mass ends $\beta$ and $\gamma$. At characteristic mass $M_1$, the ratio of stellar and subhalo mass is equal to the normalisation $N$. We fit this model to the distributions of running medians, as well as $16^\rmn{th}$ and $84^\rmn{th}$ percentiles using non-linear least squares minimisation. The fits are applied separately to the SHMRs of centrals and satellites in groups and clusters.

\section{Results}
\label{sec:res}

\subsection{Stellar-to-halo mass relation at z = 0}
\label{sec:shmr}

\begin{figure*}
    \centering
    \includegraphics[width=.6\textwidth]{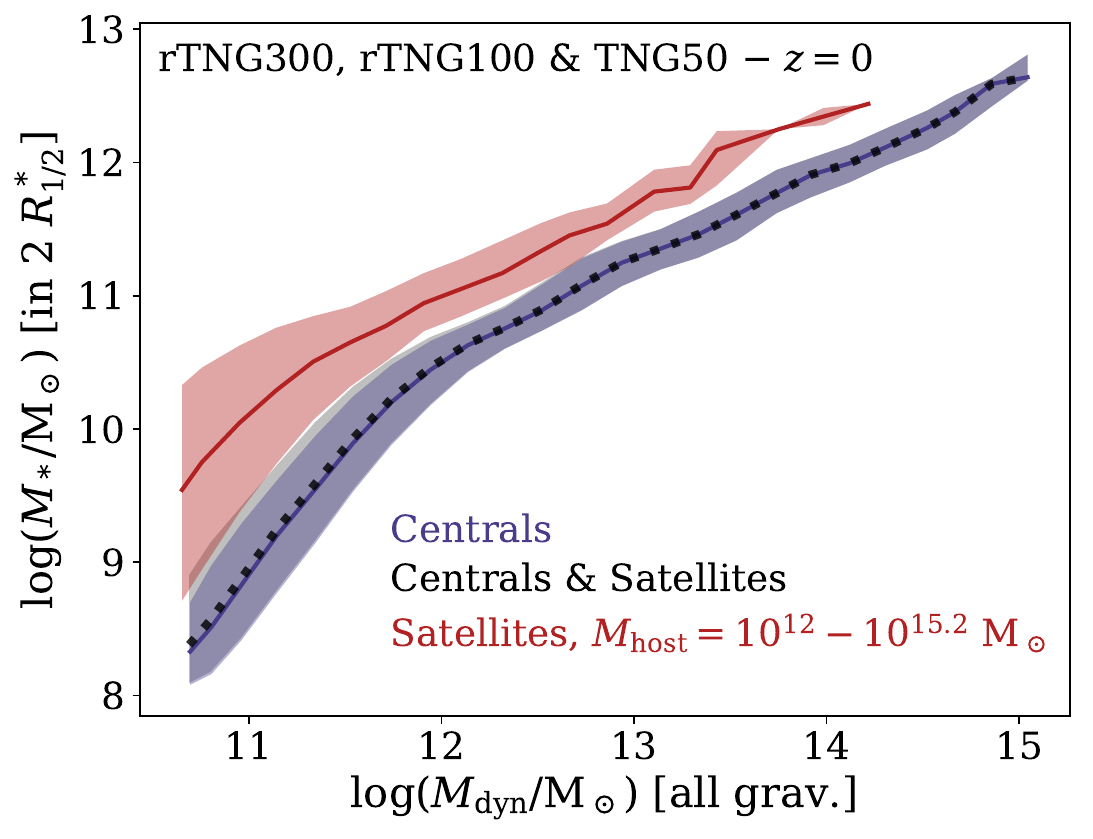}
    \includegraphics[width=\textwidth]{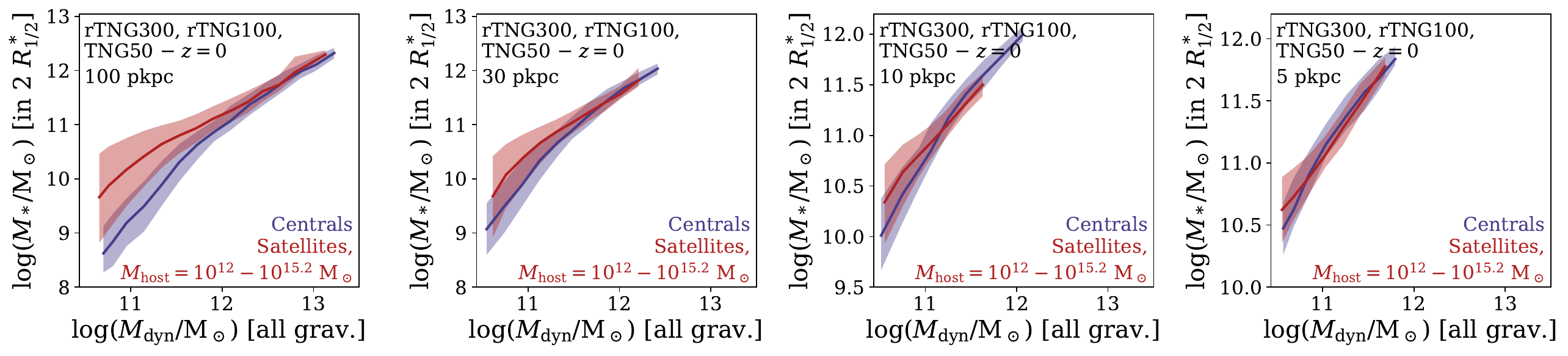}
    \caption{Stellar-to-halo mass relation for central and satellite galaxies at $z=0$ with total dynamical masses (all gravitationally bound material) of at least $10^{10.5} \text{ M}_\odot$ combining samples of the rescaled rTNG300 and rTNG100, as well as TNG50 (see text and Appendix~\ref{sec:resc} for details). In the top panel, we employ all gravitationally bound material as determined by \textsc{subfind} instead of halo mass as $M_\rmn{200c}$ for dynamical mass. Since satellites only correspond to slight enhancements on the overall background density, mass definitions using spherical overdensities would not enable a meaningful comparison between centrals and satellites. Furthermore, dynamical masses are considered \textit{not} at peak mass along each subhalo history but at present-day times, in order to highlight the impact of environmental effects. \textit{Top panel:} Stellar mass as a function of dynamical mass for centrals (solid blue curve), satellites in hosts of $10^{12} - 10^{15.2}~\rmn{M}_\odot$ (solid red curve), as well as both centrals and satellites (dotted black curve) at $z=0$, as medians within bins of $0.5~\rmn{dex}$ over the range of dynamical masses. Shaded regions correspond to $16^\rmn{th}$ and $84^\rmn{th}$ percentiles. \textit{Bottom panels:} Stellar mass as a function of dynamical mass for centrals and satellites in hosts of $10^{12} - 10^{15.2}~\rmn{M}_\odot$ in fixed physical apertures -- $100~\rmn{physical~kpc}$ (pkpc), $30~\rmn{pkpc}$, $10~\rmn{pkpc}$, $5~\rmn{pkpc}$ (from left to right).}
    \label{fig:shmr_main}
\end{figure*}

In this section, we examine the relationship of total dynamical mass $M_\rmn{dyn}$ and stellar mass $M_*$ at $z=0$, by comparing satellites in groups and clusters with $ M_\rmn{host} \geq 10^{12}~\rmn{M}_\odot$ to central galaxies. 

Figure~\ref{fig:shmr_main} shows the SHMR of galaxies with $M_\rmn{dyn} \geq 10^{10.5}~\rmn{M}_\odot$ in TNG50 and the resolution-rescaled rTNG100 and rTNG300 (see Appendix~\ref{sec:resc}): centrals (solid blue curve), satellites (solid red curve), as well as both centrals and satellites combined (dotted black curve). We consider masses in our fiducial aperture choice -- the sum of all gravitationally bound particles for total dynamical mass and all stellar particles within twice the stellar half-mass radius $R_{1/2}^*$ for stellar mass. There is a systematic offset between central and satellite galaxy populations: at fixed stellar mass, satellites are shifted towards smaller total dynamical mass. Shaded areas show the scatter in the SHMR as $16^\rmn{th}$ and $84^\rmn{th}$ percentiles. At all dynamical masses, satellites exhibit a larger scatter than centrals, increasing towards the lower mass end.

Additionally, we present combinations of fixed physical apertures in the bottom panels. Here, both stellar and subhalo mass are confined to the innermost $100~\rmn{pkpc}$ (physical kpc), $50~\rmn{pkpc}$, $10~\rmn{pkpc}$ and $5~\rmn{pkpc}$ (from left to right). Measuring stellar and dynamical masses within fixed physical apertures shows a similar offset for the largest aperture of $100~\rmn{pkpc}$. However, the offset between satellites and centrals at the high-mass end is less pronounced than for our fiducial apertures. While $100~\rmn{pkpc}$ still encompass all gravitationally bound particles in low- and intermediate-mass subhaloes, the upper limit of dynamical mass shifts to a lower value compared to the SHMR in our fiducial aperture choice. Since the dark matter subhalo is more extended than the galaxy's stellar body, this affects the total dynamical mass to a larger degree than the stellar mass. When the SHMR is examined for progressively smaller apertures, the offset between centrals and satellites becomes less significant over the whole range of dynamical mass, albeit to a lesser degree towards the low-mass end for larger apertures. Environmental effects that cause this offset between the SHMRs of centrals and satellites affect galaxies in an outside-in fashion. Since the inner galaxy regions remain largely unaffected by their environment, the offset between the SHMRs of centrals and satellites decreases when constraining galaxy and subhalo mass to smaller apertures.

\subsection{Dependence on host mass and redshift}

\begin{figure*}
    \centering
    \includegraphics[width=.45\textwidth]{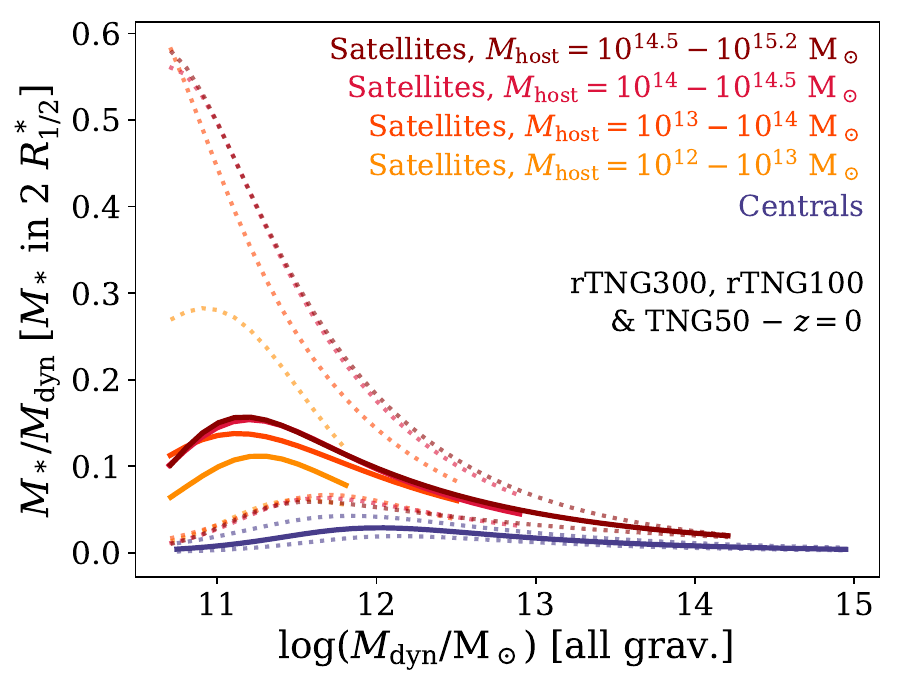} \hspace{.3cm}
    \includegraphics[width=.45\textwidth]{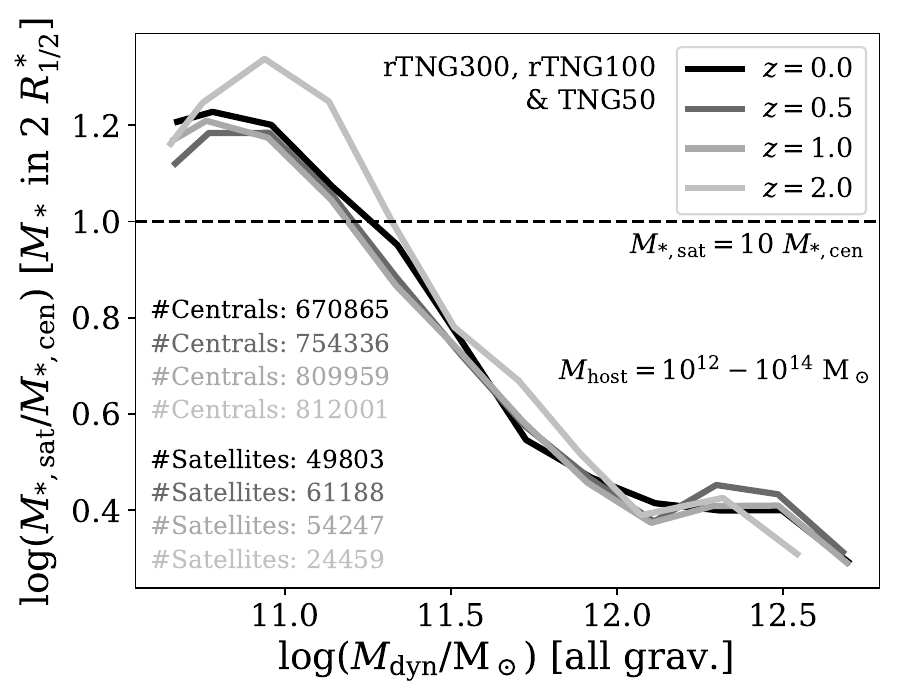}
    \caption{Stellar-to-halo mass relation for central and satellite galaxies with total dynamical masses of at least $10^{10.5} \text{ M}_\odot$ from rTNG300, rTNG100, and TNG50 as a function of host mass and redshift. \textit{Left panel:} SHMR for centrals and subsamples of satellites within fiducial apertures at $z=0$. Satellites are divided by host mass into bins of $10^{12} - 10^{13}~\rmn{M}_\odot$, $10^{13} - 10^{14}~\rmn{M}_\odot$, $10^{14} - 10^{14.5}~\rmn{M}_\odot$, and $10^{14.5} - 10^{15.2}~\rmn{M}_\odot$ (orange to dark red, solid curves). The most massive host mass bin only includes rTNG300 galaxies, while the others combine galaxies from rTNG300, rTNG100 and TNG50. Relations are shown as fits to the running medians of stellar mass fractions $M_*/M_\rmn{dyn}$ within bins of $0.7~\rmn{dex}$ (solid curves). Dotted curves correspond to fits to their $16^\rmn{th}$ and $84^\rmn{th}$ percentiles. \textit{Right panel:} Stellar mass ratios of satellite to central galaxies in rTNG300, rTNG100 and TNG50 as a function of dynamical mass at $z=0.0, 0.5, 1.0, 2.0$ (black to light grey curves). We limit satellites to hosts of $10^{12} - 10^{14}~\rmn{M}_\odot$, since TNG50 does not include $>10^{14}~\rmn{M}_\odot$ haloes at earlier redshifts.}
    \label{fig:shmr_main2}
\end{figure*}

We examine the separation of satellite galaxies more closely in the left panel of Figure~\ref{fig:shmr_main2}. Here, satellite galaxies of rTNG300, rTNG100, and TNG50 are split into subsamples according to their $z=0$ host mass: $10^{12} - 10^{13}~\rmn{M}_\odot$, $10^{13} - 10^{14}~\rmn{M}_\odot$, $10^{14} - 10^{14.5}~\rm{M}_\odot$, and $10^{14.5} - 10^{15.2}~\rm{M}_\odot$. The most massive host mass bin includes exclusively rTNG300 satellites, while the other three bins consist of satellites from rTNG300, rTNG100, and TNG50.

The SHMR is shown as fits to the average distribution of stellar mass fractions at a given dynamical mass for centrals and the four satellite subsamples, following the fitting function in Section~\ref{sec:fit}. We fit Equation~\eqref{eq:mosterfit} to the distributions of running medians (solid curves), as well as $16^\rmn{th}$ and $84^\rmn{th}$ percentiles (dotted curves) to depict the differences in scatter between centrals and satellites in groups and clusters. 

The SHMR of satellite galaxies generally shows a large offset from the SHMR of centrals, with satellite subhaloes exhibiting larger stellar mass fractions over the whole range of dynamical mass. We quantify this offset at the peak of the relation, ranging from stellar-to-halo mass ratios of about 10 per cent for satellites in $10^{12} - 10^{13}~\rmn{M}_\odot$ hosts to 15 per cent in hosts of $10^{14.5} - 10^{15.2}~\rmn{M}_\odot$. 

While there is a trend with host mass -- satellites in more massive hosts tend to have in the median larger stellar mass fractions at fixed dynamical mass -- this correlation is even more pronounced when considering the relation's scatter. While the distribution of $16^\rmn{th}$ percentiles practically shows the same basic offset from the SHMR of centrals for all satellites, the $84^\rmn{th}$ percentiles of SHMRs increase more significantly than the average median relation. Satellites in more massive environments can reach larger stellar mass fractions: up to 28 per cent in hosts of $10^{12} - 10^{13}~\rmn{M}_\odot$ or 50--60 per cent in hosts of $10^{13} - 10^{15.2}~\rmn{M}_\odot$ at the peak of $84^\rmn{th}$ percentiles. On the other hand, the maximum stellar mass fraction for the $84^\rmn{th}$ percentiles of central galaxies only reaches 2--4 per cent. The fit parameters of the four samples' average distributions are summarised in Table~\ref{tab:fits_censat}. Furthermore, the same trends hold for general baryonic-to-total mass ratios considering the contributions of both stars and gas. We emphasise that the SHMRs for satellites in $10^{14} - 10^{14.5}~\rmn{M}_\odot$ and $10^{14.5} - 10^{15.2}~\rmn{M}_\odot$ hosts represent lower limits due to effects of numerical resolution: the rescaling process for stellar masses of satellites in hosts of $10^{14} - 10^{15.2}~\rmn{M}_\odot$ relies on only one massive cluster in TNG50 with a mass of $10^{14.3}~\rmn{M}_\odot$. Therefore, the SHMRs of satellites within hosts of this mass range may in reality be shifted to even larger stellar mass fractions.

In the right panel of Figure~\ref{fig:shmr_main2}, we show the stellar mass ratio of satellites and centrals in rTNG300, rTNG100, and TNG50 as a function of total dynamical mass and its evolution with time. However, we only consider satellites in hosts of $10^{12} - 10^{14}~\rmn{M}_\odot$, since TNG50 does not include $10^{14}~\rmn{M}_\odot$ haloes at $z=0.5$ and earlier redshifts. At all redshifts considered ($z=0,0.5,1,2$; black to light grey curves), the samples include tens of thousands of satellites and hundreds of thousands of centrals. At fixed dynamical mass, the stellar mass of satellites exhibits a significant difference to those of centrals -- larger by a factor of at least 2.5 at $z=0$. This increases substantially for subhaloes with $M_\rmn{dyn} < 10^{12}~\rmn{M}_\odot$ -- around which satellite subhaloes reach peak baryonic conversion efficiency -- and reaches its maximum at our lower dynamical mass limit of $10^{10.5}~\rmn{M}_\odot$. Here, satellites are more massive in stars than centrals by a factor of 16 at $z=0$, $z=0.5$, and $z=1$, as well as a factor of 22 at $z=2$. However, there is no statistically significant difference in the ratios of stellar mass between satellites and centrals from $z=0$ to $z=2$. Satellites already exhibit an offset in stellar mass at fixed dynamical mass as compared to those of centrals at early times: since the density profiles of both satellites and host environments stay on average similar between $z=2$ and $z=0$, tidal stripping in the host halo's gravitational potential operates -- for satellites of a given dynamical mass -- to the same degree at different redshifts.

\begin{table*}
    \centering
    \begin{tabular}{l c c c c}
        \hline \hline
        Sample & $N$ & $M_1 [\log \rmn{M}_\odot]$ & $\beta$ & $\gamma$\\ \hline
        Centrals & $0.0258 \pm 0.0003$ & $11.70 \pm 0.02$ & $28.6 \pm 0.8$ & $10.4 \pm 0.2$\\
        Satellites in $10^{12} - 10^{13}~\rmn{M}_\odot$ hosts & $0.108 \pm 0.003$ & $11.12 \pm 0.06$ & $27.5 \pm 2.9$ & $15.6 \pm 1.7$\\
        Satellites in $10^{13} - 10^{14}~\rmn{M}_\odot$ hosts & $0.127 \pm 0.008$ & $10.85 \pm 0.11$ & $23.6 \pm 6.1$ & $10.1 \pm 1.3$\\
        Satellites in $10^{14} - 10^{14.5}~\rmn{M}_\odot$ hosts & $0.137 \pm 0.004$ & $10.93 \pm 0.04$ & $30.5 \pm 3.6$ & $10.9 \pm 0.6$\\
        Satellites in $10^{14.5} - 10^{15.2}~\rmn{M}_\odot$ hosts & $0.129 \pm 0.006$ & $10.85 \pm 0.04$ & $38.6 \pm 7.2$ & $9.5 \pm 0.7$\\\hline
    \end{tabular}
    \caption{Fit parameters for the SHMR of centrals and satellites as a function of host mass in the left panel of Figure~\ref{fig:shmr_main2} using rTNG300, rTNG100, and TNG50. We follow the parametrization in Equation~\eqref{eq:mosterfit} \protect\citep[similar to][]{moster2010,moster2013}: \text{normalisation $N$}, characteristic mass $M_1$, and the slopes at the low- and high-mass ends $\beta$ and $\gamma$.}
    \label{tab:fits_censat}
\end{table*}

\subsection{Scatter in the stellar-to-halo mass relation}
\label{sec:mstar_scatter}

\begin{figure*}
\centering
\includegraphics[width=.6\textwidth]{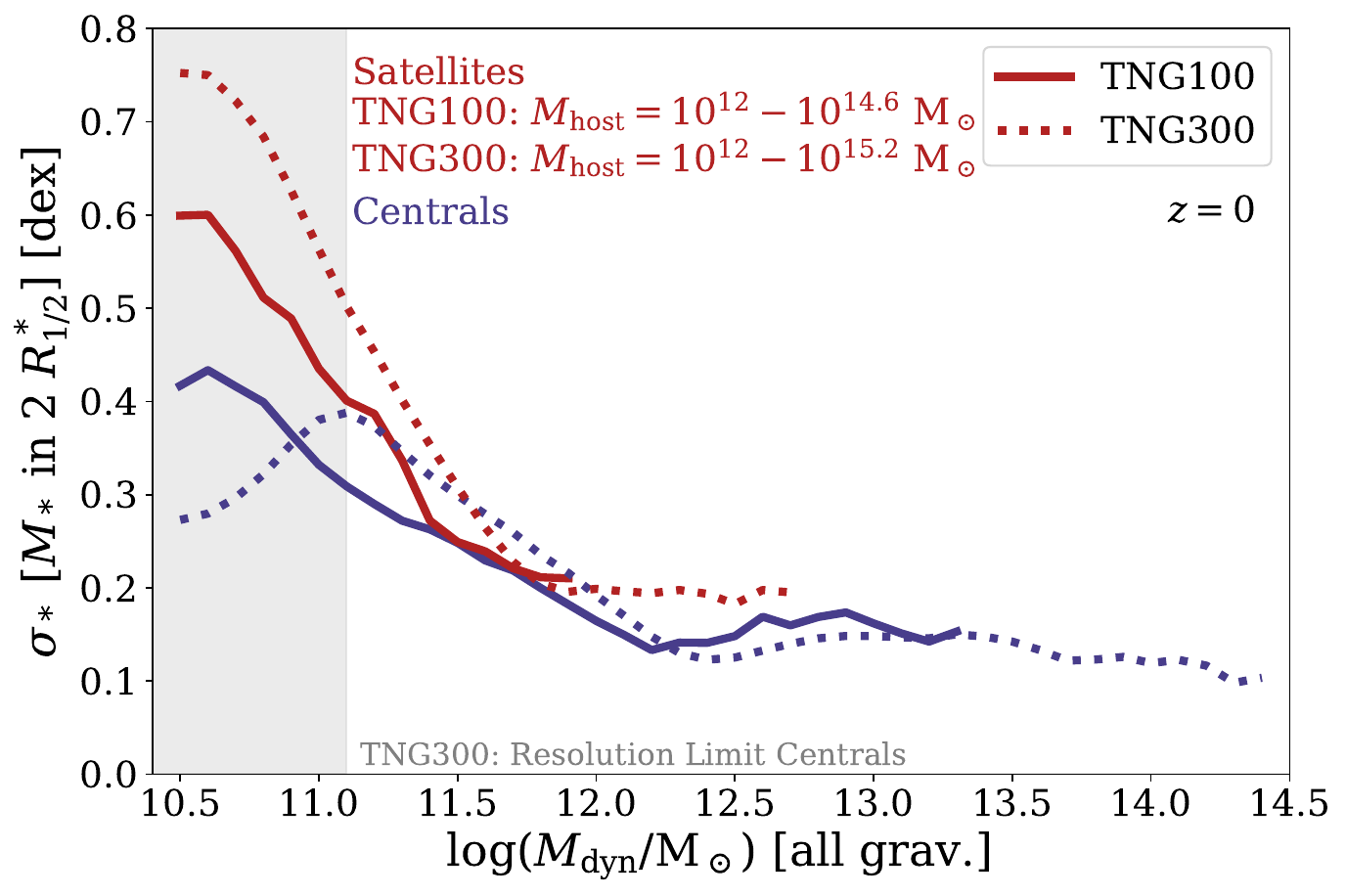}
\includegraphics[width=.45\textwidth]{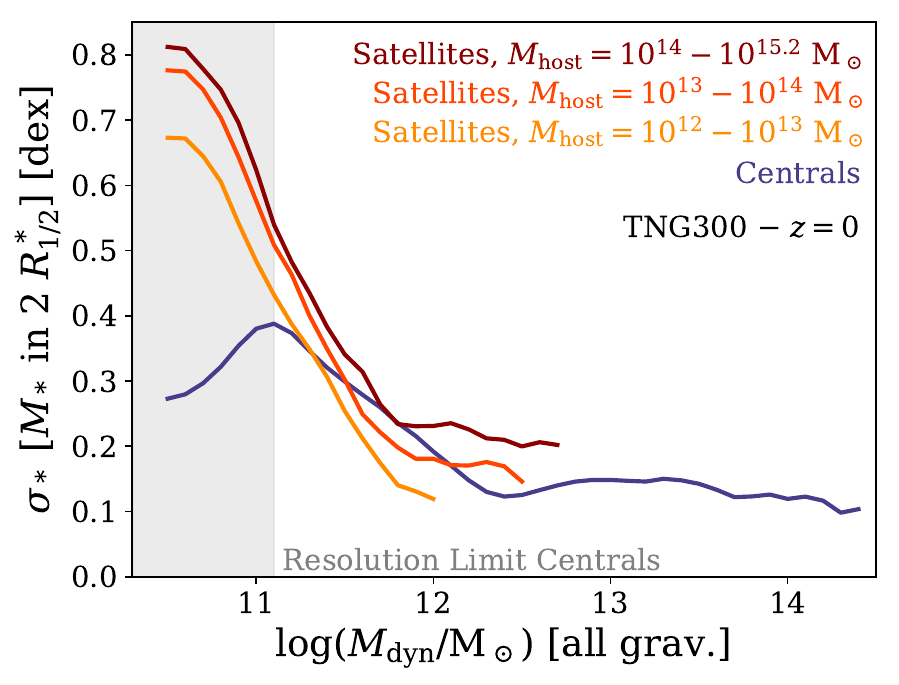} \hspace{.5cm}
\includegraphics[width=.45\textwidth]{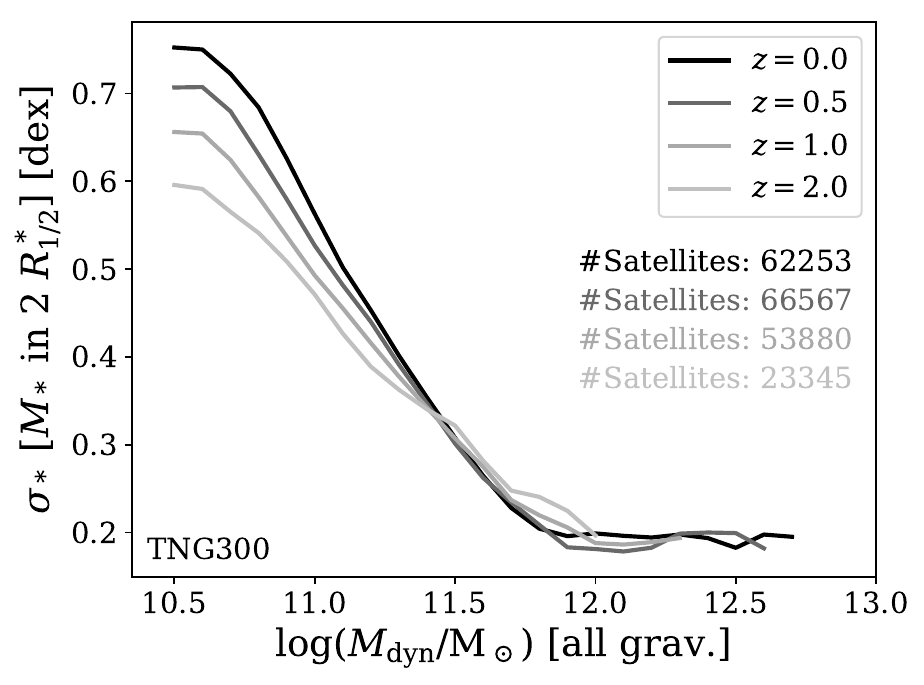}
\caption{\textit{Top panel:} Scatter in (logarithmic) stellar mass, $\sigma_*$, as a function of total dynamical mass for centrals (blue curves) and satellites (red curves) in TNG100 (solid curves) and TNG300 (dotted curves). The grey area denotes the resolution limit where the sample of centrals in TNG300 includes galaxies with only a single stellar particle and the distribution of stellar mass is no longer fully sampled. \textit{Bottom left panel:} Stellar mass scatter $\sigma_*$ as a function of total dynamical mass in TNG300 for centrals (blue curve) and satellites in different bins of host mass (orange to red curves). \textit{Bottom right panel:} Scatter $\sigma_*$ of satellites in hosts of at least $10^{12}~\rmn{M_\odot}$ as a function of dynamical satellite mass at different redshifts: $z=0.0, 0.5, 1.0, 2.0$.}
\label{fig:mstar_scatter_TNG100-300}
\end{figure*}

The environment affects the dark matter subhalo and the stellar body of a galaxy to a different degree, which results not only in an offset between centrals and satellites in groups and clusters in the SHMR but also in different scatter along the relation. In this section, we examine the scatter in stellar mass $\sigma_*$ as a function of total dynamical mass and the ways in which the environment shapes it. We determine the stellar mass scatter by defining bins of fixed dynamical mass and by computing the standard deviation of the distribution of logarithmic stellar mass within. These distributions correspond approximately to Gaussians \citep[for non-logarithmic masses this corresponds to a lognormal distribution, see also][]{anbajagane2020}.

The top panel of Figure~\ref{fig:mstar_scatter_TNG100-300} shows the scatter as a function of total dynamical mass for all centrals (blue curves) and satellites (red curves) in hosts of $10^{12} - 10^{14.6}~\rm{M}_\odot$ in TNG100 (solid curves) as well as hosts of $10^{12} - 10^{15.2}~\rm{M}_\odot$ in TNG300 (dotted curves). However, the low-mass end of TNG300 centrals reaches the resolution limit (grey area): here, our sample of centrals starts to include galaxies with only a single stellar particle and the SHMR's scatter is no longer fully sampled. Since the distribution of stellar mass within fixed dynamical mass bins is incomplete the scatter decreases. In both simulations there is a significant offset between centrals and satellites in groups and clusters. The scatter of centrals and satellites increases towards lower dynamical masses to up to $0.43~\rmn{dex}$ for centrals and $0.60~\rmn{dex}$ for satellites in TNG100, as well as $0.38~\rmn{dex}$ for centrals and $0.77~\rmn{dex}$ for satellites in TNG300. Considered at the respective peak scatter of centrals, this results in an offset of $0.17~\rmn{dex}$ at $M_\rmn{dyn} = 10^{10.6}~\rmn{M}_\odot$ for satellites in TNG100 and $0.12~\rmn{dex}$ at $M_\rmn{dyn} = 10^{11.1}~\rmn{M}_\odot$ for satellites in TNG300. For TNG100, this dynamical mass yields an offset of only $0.1~\rmn{dex}$. 

As galaxies become less massive, the scatter increases for both centrals and satellites. While this effect is mainly driven by different assembly histories for centrals, it is even more pronounced for low-mass satellites as they become less resistant to their environment. For intermediate- to high-mass subhaloes ($M_\rmn{dyn} \gtrsim 10^{12}~\rmn{M}_\odot$ for centrals, $M_\rmn{dyn} \gtrsim 10^{11.5}~\rmn{M}_\odot$ for satellites) the scatter becomes constant around a value of $\sigma_* \sim 0.2~\rmn{dex}$ for satellites and $\sigma_* \sim 0.15~\rmn{dex}$ for centrals in both TNG100 and TNG300. For both centrals and satellites, constant scatter sets in for subhaloes that correspond to the SHMR's peak -- subhaloes of peak star formation efficiency -- and continues to their respective high mass ends. 

We examine the effects of group and cluster environments separately in the bottom left panel of Figure~\ref{fig:mstar_scatter_TNG100-300} by splitting satellite galaxies in TNG300 by host mass. Over the whole range of dynamical mass, there is a continuous offset between satellites in different hosts. Satellites in hosts of $10^{14} - 10^{15.2}~\rmn{M}_\odot$ and $10^{13} - 10^{14}~\rmn{M}_\odot$ show the largest scatter of up to $0.8~\rmn{dex}$, while satellites in $10^{12} - 10^{13}~\rmn{M}_\odot$ hosts reach up to $0.7~\rmn{dex}$. However, even satellites in less massive hosts already exhibit a significant difference to the centrals' relation. Considered at a dynamical mass of $10^{11.1}~\rmn{M}_\odot$ -- corresponding to the peak scatter of centrals -- satellites show an offset of $0.15~\rmn{dex}$, $0.12~\rmn{dex}$, and $0.05~\rmn{dex}$ (in decreasing host mass bins) compared to centrals of the same mass. For all satellites, the scatter in stellar mass becomes constant around their respective subhalo mass of peak star formation efficiency. The offset between satellites in more and less massive hosts remains constant at the high subhalo mass end with satellites in $10^{12} - 10^{13}~\rmn{M}_\odot$ hosts settling around a scatter of $0.14~\rmn{dex}$ -- similar to the scatter of centrals.

The lower right panel of Figure~\ref{fig:mstar_scatter_TNG100-300} shows the evolution in time of the scatter $\sigma_*$ for satellite galaxies in TNG300. This includes several tens of thousands of satellites at the redshifts considered ($z=0,0.5,1,2$). At all redshifts, the scatter of satellites at the massive dynamical mass end with $M_\rmn{dyn} \gtrsim 10^{12}~\rmn{M}_\odot$ is roughly constant at $\sigma_* \sim 0.2~\rmn{dex}$. However, for lower-mass satellites, there is a slight, albeit clear trend of decreasing scatter with increasing redshift: while the scatter reaches up to $0.77~\rmn{dex}$ at $z=0$, this peak value decreases continuously to $0.72~\rmn{dex}$ at $z=0.5$, $0.67~\rmn{dex}$ at $z=1$, and $0.61~\rmn{dex}$ at $z=2$. Although our satellite sample shows no trend in its average SHMR at different times (see Figure~\ref{fig:shmr_main}), the scatter of stellar mass at fixed dynamical mass builds up over time. The scatter in the SHMR of centrals, on the other hand, only shows a slight increase in scatter with increasing redshift, consistent with \cite{pillepich2018b}.

\subsection{Dependence on environment and accretion history}
\label{sec:shmr_env}

\begin{figure}
\centering
\includegraphics[width=.45\textwidth]{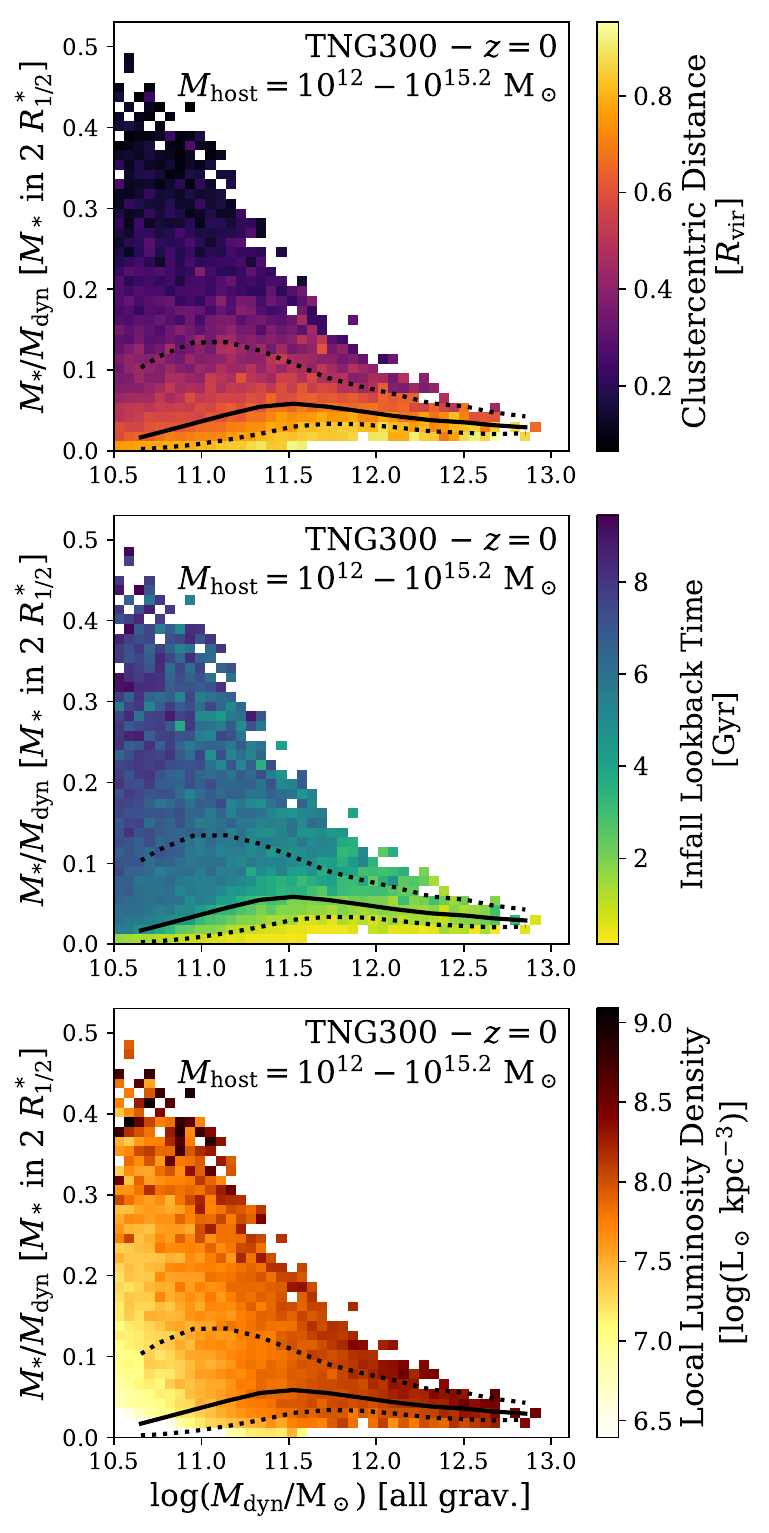}
\caption{Stellar-to-halo mass relation of satellite galaxies in hosts of at least $10^{12}~\rmn{M}_\odot$ at $z=0$ in TNG300. We divide the SHMR into 2D bins and colour-code bins that contain at least five satellites by their median value of cluster-centric distance (top panel), lookback time to the first infall into the virial radius of the satellites' present-day host halo (middle panel), and local luminosity density for galaxies within $0.1~R_\rmn{vir}$ (bottom panel). Solid black curves show the medians in bins of total dynamical mass with width $0.5~\rmn{dex}$, dotted curves correspond to $16^\rmn{th}$ and $84^\rmn{th}$ percentiles.}
\label{fig:shmr_env_allsats}
\end{figure}

\begin{figure*}
\centering
\includegraphics[width=.95\textwidth]{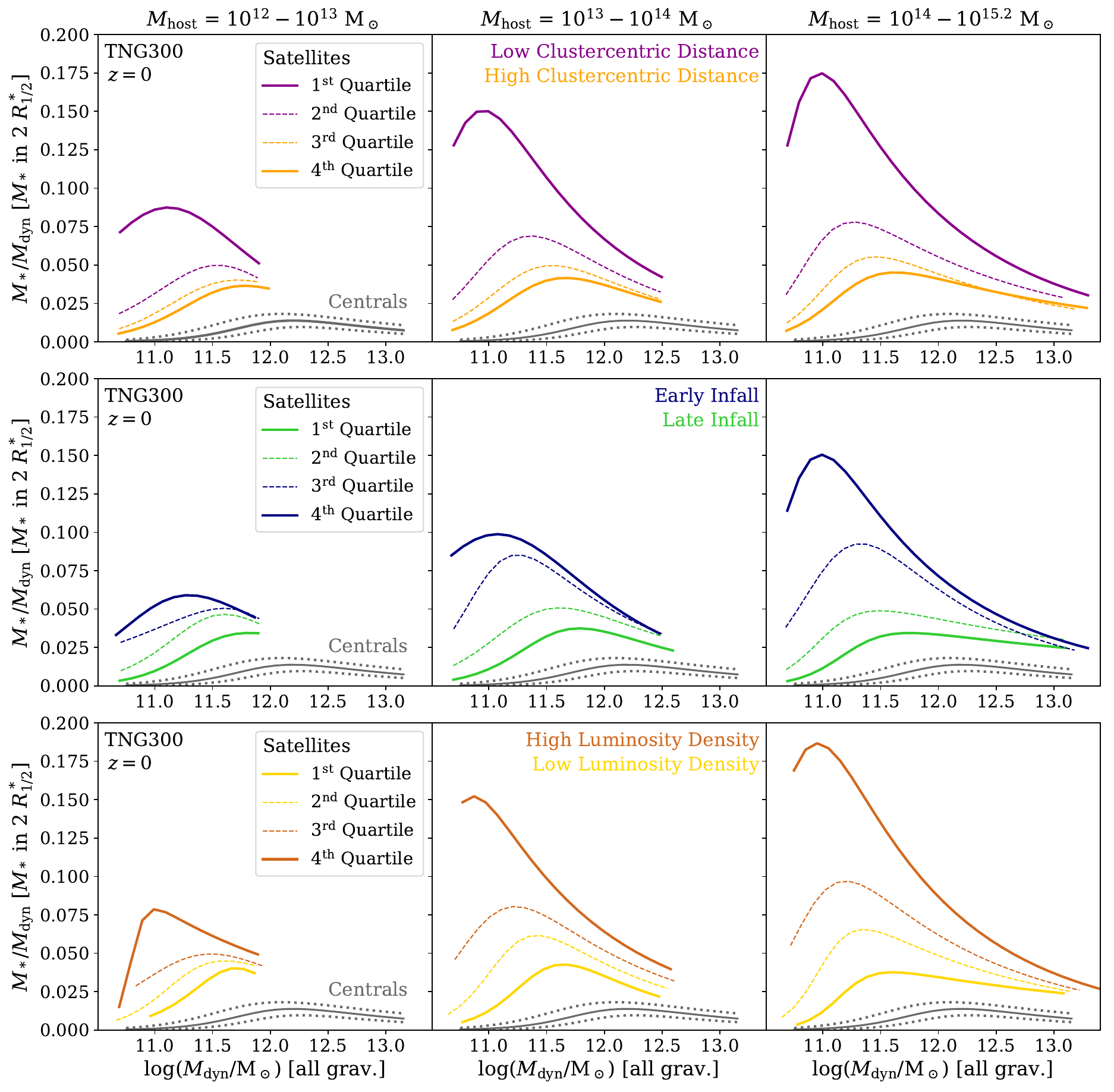}
\caption{Stellar-to-halo mass relation as a function of environment and host mass in TNG300 at $z=0$. Each row shows an environmental parameter  (\textit{from top to bottom:} cluster-centric distance, infall time into the current host's virial radius, and local luminosity density). Each column corresponds to a different host halo mass range (\textit{from left to right:} $10^{12} - 10^{13}~\rmn{M}_\odot$, $10^{13} - 10^{14}~\rmn{M}_\odot$, $10^{14} - 10^{15.2}~\rmn{M}_\odot$). Satellite galaxies are split into quartiles of low/high cluster-centric distance (purple/orange), early/late infall (blue/green), as well as low/high local luminosity density (yellow/brown) within bins of total dynamical mass with width $0.7~\rmn{dex}$. Solid curves correspond to first and fourth quartiles, dashed curves to second and third quartiles. Grey lines depict the SHMR of central galaxies as fits to the moving averages (solid curves) as well as $16^\rmn{th}$ and $84^\rmn{th}$ percentiles (dotted curves) to account for scatter in the relation.}
\label{fig:shmr_env_offset}
\end{figure*}

In this section, we investigate the connection of satellites and their environment more closely. Since host mass is not the only property that describes galaxy environment, we employ cluster-centric distance to account for the varying strength of cluster potentials, infall times to account for the period over which satellites have been exposed to environmental influence, and a local luminosity density to account for the immediate surroundings of satellites. Infall times correspond to the first time satellites crossed the virial radius of their present-day host halo's main progenitor (see Section~\ref{sec:sample} for details).

Figure~\ref{fig:shmr_env_allsats} illustrates the SHMR of satellites in hosts of at least $10^{12}~\rmn{M}_\odot$ as a function of said environmental properties in TNG300. Bins including at least five satellites are colour-coded by their respective median values of cluster-centric distance (top panel), time of infall into their present-day host's virial radius (middle panel), and local luminosity density (bottom panel). Here we show results from TNG300 (without resolution correction) as we are focusing on relative effects.

At fixed dynamical mass, galaxies with larger stellar mass fractions reside on average closer to the cluster centre (where the host halo's gravitational potential is deeper), experienced an early infall into the virial radius of their present-day host, and are located in areas of higher local density. Lower stellar mass fraction satellites, on the other hand, reside at higher cluster-centric distances, fell later into their present-day environment, and inhabit regions of lower density. They have been exposed to weaker environmental effects for a shorter amount of time -- and are closer to the distribution of central galaxies in the SHMR. However, there is an additional bias with dynamical mass for local luminosity density since more massive subhaloes host more luminous objects. At the high dynamical mass end, the correlation of stellar mass fractions with local density becomes less pronounced. Black curves correspond to the average SHMR (solid curves) as well as to the $16^\rmn{th}$ and $84^\rmn{th}$ percentiles (dotted curves) of the satellites. Only a small fraction of satellites contributes to the high stellar mass fraction tail, which can reach up to 50 per cent at the low dynamical mass end.

We quantify the differences for satellite subpopulations in Figure~\ref{fig:shmr_env_offset} and show the SHMR as a function of environment for TNG300 satellites at $z=0$ in three bins of host mass: $10^{12} - 10^{13}~\rmn{M}_\odot$, $10^{13} - 10^{14}~\rmn{M}_\odot$, and $10^{14} - 10^{15.2}~\rmn{M}_\odot$ (from left to right). At a given dynamical mass, we divide the satellites into four quartiles with respect to each environmental quantity and fit the model in Equation~\eqref{eq:mosterfit} to the resulting SHMRs. Thus we are able to examine the relations of low and high cluster-centric distance populations (magenta/orange curves), early and late infallers (depending on host mass with respect to 2.5--4 Gyr ago; blue/green curves), as well as satellites in low and high luminosity density environments separately (yellow/brown curves). Furthermore, we include the average SHMR of centrals (solid grey curves), as well as their $16^\rmn{th}$ and $84^\rmn{th}$ percentiles (dotted grey curves). 

Clearly, the SHMR of satellite galaxies correlates with their environment, with the overall scatter and the offsets of the respective quartiles (low cluster-centric distance, early infall, high local luminosity density) increasing significantly with host mass. For all hosts and all environmental parameters, even the satellite subsamples that are subject to a weaker influence by their environment (i.e. high cluster-centric distance, late infall, low local density) already feature a significant offset from the centrals' SHMR. Peak stellar mass fractions range from 3 per cent for late-infall satellites in both $10^{14} - 10^{15.2}~\rmn{M}_\odot$ and $10^{12} - 10^{13}~\rmn{M}_\odot$ hosts to 4 per cent for satellites in low luminosity density areas of $10^{13} - 10^{14}~\rmn{M}_\odot$ hosts. On the other hand, satellites that have been subject to stronger environmental effects (i.e. low cluster-centric distance, early infall, high local density) clearly exhibit even larger offsets from the SHMR of centrals, increasing with host mass. Their SHMRs reach peak stellar mass fractions ranging from 6 per cent for early infallers in $10^{12} - 10^{13}~\rmn{M}_\odot$ hosts to up to 18 per cent for satellites in high luminosity density regions of $10^{14} - 10^{15.2}~\rmn{M}_\odot$ hosts. While local luminosity density serves as a reasonable estimate of environmental impact in massive clusters of $10^{14} - 10^{15.2}~\rmn{M}_\odot$, these trends appear less regular in lower mass groups and more sparsely populated environments.

\begin{figure*}
    \centering
    \includegraphics[width=.95\textwidth]{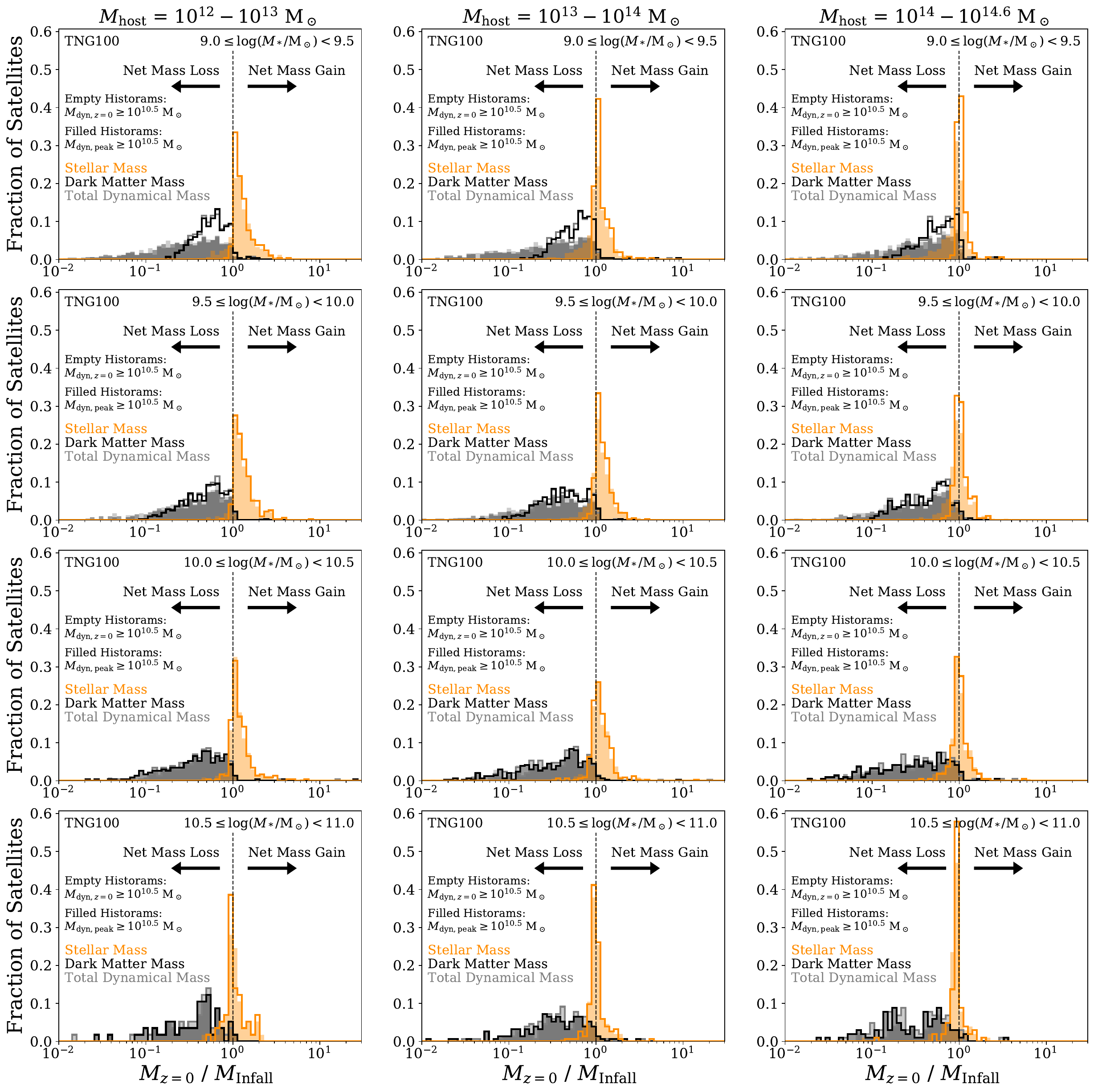}
    \caption{Ratios of satellite mass between $z=0$ and first infall for satellite galaxies in TNG100. We define infall as the first time a satellite crosses the virial radius $R_\rmn{200c}$ of its present-day host. Mass ratios are shown for stellar (orange), dark matter (black) and total dynamical mass (grey) in our fiducial aperture choice: all gravitationally bound particles for total dynamical and dark matter mass as well as all stellar particles within two stellar half-mass radii for stellar mass. We show the distributions as a function of host mass across columns and satellite stellar mass across rows: $10^{12} - 10^{13}~\rmn{M}_\odot$, $10^{13} - 10^{14}~\rmn{M}_\odot$, and $10^{14} - 10^{14.6}~\rmn{M}_\odot$ in host mass (from left to right), as well as $10^{9} - 10^{9.5}~\rmn{M}_\odot$, $10^{9.5} - 10^{10}~\rmn{M}_\odot$, $10^{10} - 10^{10.5}~\rmn{M}_\odot$, and $10^{10.5} - 10^{11}~\rmn{M}_\odot$ in satellite stellar mass (from top to bottom). In addition to our fiducial satellite selection with present-day dynamical mass of $M_{\rmn{dyn, }z=0} \geq 10^{10.5}~\rmn{M}_\odot$ (empty histograms), we show the mass ratios for all surviving satellites that reached a peak dynamical mass of $M_\rmn{dyn, peak} \geq 10^{10.5}~\rmn{M}_\odot$ at some point in their lifetime (filled histograms).}
    \label{fig:massratios}
\end{figure*}

\begin{figure*}
    \centering
    \includegraphics[width=.95\textwidth]{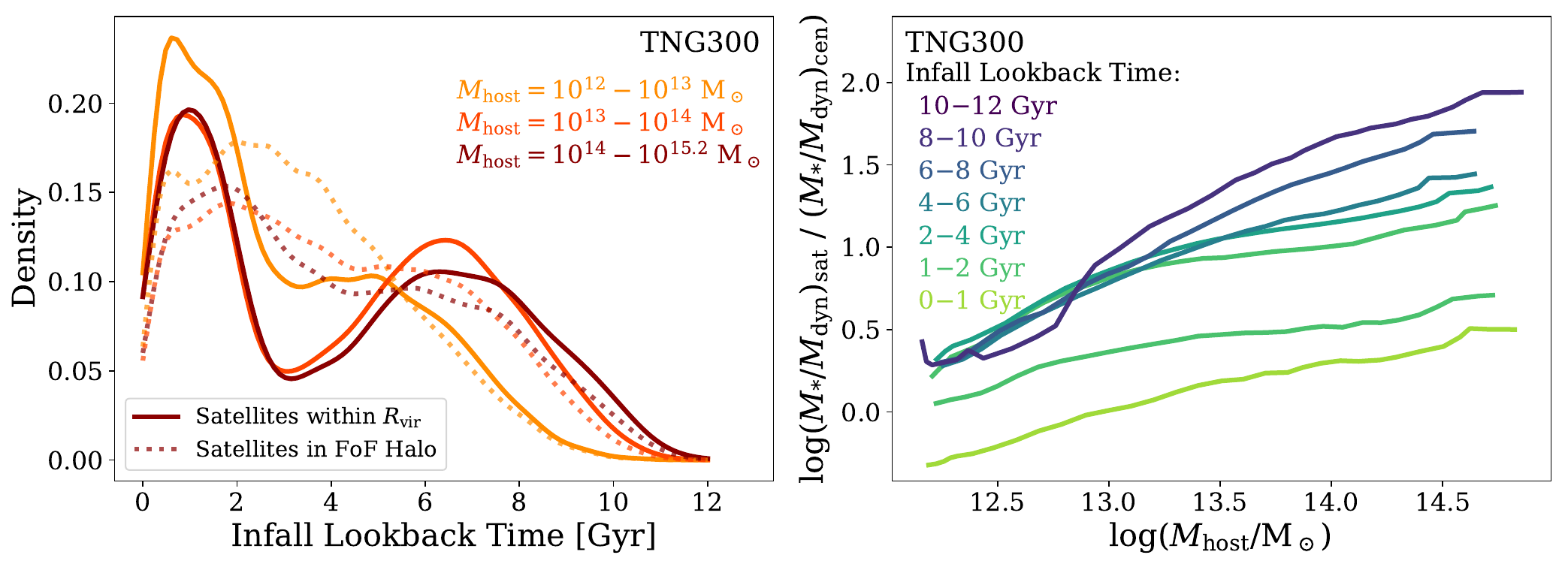}
    \caption{\textit{Left panel:} Distribution of infall times for satellite galaxies in TNG300. We present their accretion history as a function of host mass in bins of $10^{12} - 10^{13}~\rmn{M}_\odot$, $10^{13} - 10^{14}~\rmn{M}_\odot$, and $10^{14} - 10^{15.2}~\rmn{M}_\odot$ (orange to dark red curves). Solid curves correspond to satellite galaxies within their host's virial radius, dotted curves to all satellites in the host's FoF halo, i.e. within \textit{and} outside the virial radius. Infall distributions were smoothed using a Gaussian kernel with an average width of 0.3 Gyr. \textit{Right panel:} Ratios of satellite and central stellar mass fractions $M_*/M_\rmn{dyn}$ as a function of host mass in TNG300. We show the relation in different bins of infall lookback time: 0--1 Gyr, 1--2 Gyr, 2--4 Gyr, 4--6 Gyr, 6--8 Gyr, 8--10 Gyr, and 10--12 Gyr ago (green to blue curves).}
    \label{fig:hostmass_infall}
\end{figure*}

\section{Interpretation, tools, and discussion}
\label{sec:disc}

\subsection{Transition of satellite galaxies: tidal mass loss vs. quenching}
\label{sec:massloss}

We can attribute the offset between the SHMRs of centrals and satellites for the most part to tidal stripping of satellites in interactions with the host halo's gravitational potential and the loss of their dark matter subhalo. Figure~\ref{fig:massratios} illustrates the effects of environment on total dynamical mass as well as the stellar and dark matter components of TNG100 satellites over time. We show the ratio of their masses between $z = 0$ and the first infall into the virial radius of their present-day host's main progenitor for stellar (orange), dark matter (black), and total dynamical mass (grey) in our fiducial aperture choice (all gravitationally bound particles for $M_\rmn{dyn}$ and $M_\rmn{DM}$, all stellar particles within two stellar half-mass radii for $M_*$). Furthermore, satellites are divided by the mass of their host into bins of $10^{12} - 10^{13}~\rmn{M}_\odot$, $10^{13} - 10^{14}~\rmn{M}_\odot$, and $10^{14} - 10^{14.6}~\rmn{M}_\odot$ (increasing from left to right), as well as divided by their stellar mass in bins of $10^9 - 10^{9.5}~\rmn{M}_\odot$, $10^{9.5} - 10^{10}~\rmn{M}_\odot$, $10^{10} - 10^{10.5}~\rmn{M}_\odot$, and $10^{10.5} - 10^{11}~\rmn{M}_\odot$ (from top to bottom). In this Figure, we illustrate the mass ratios for two different samples of satellite galaxies: our fiducial satellite selection with present-day dynamical mass of $M_{\rmn{dyn, }z=0} \geq 10^{10.5}~\rmn{M}_\odot$ (empty histograms), and satellites that reached a peak dynamical mass of $M_\rmn{dyn, peak} \geq 10^{10.5}~\rmn{M}_\odot$ at some point throughout their lifetime (filled histograms). So the latter sample additionally includes satellite galaxies with present-day dynamical masses of less than $10^{10.5}~\rmn{M}_\odot$.

For the most part, the mass ratios of dark matter and total dynamical mass coincide with each other. Their distributions show almost exclusively mass ratios smaller than unity, corresponding to a net mass loss due to tidal stripping of the satellites' dark matter subhaloes -- regardless of stellar mass or host mass bins. While it appears as if galaxies of larger stellar mass are subject to a stronger degree of tidal stripping of dark matter and total mass for our fiducial sample in the empty histograms, the higher mass loss tails are actually restricted by our initial subhalo selection of $M_\rmn{dyn} \geq 10^{10.5}~\rmn{M}_\odot$. The tidal mass loss tails of our alternative sample in the filled histograms, which include less massive satellites, all have a similar extent irrespective of satellite stellar mass. For larger satellite stellar masses in the bottom panels, the distributions of dark matter and dynamical mass ratios of both satellite samples coincide with each other. In these cases, tidal stripping did not put satellites in the original selection below our selection limit.

The evolution of the satellites' stellar mass component after infall is dominated by star formation. Most satellites show a net mass gain in stellar mass with mass ratios greater than unity. However, satellites in the most massive stellar mass bin exhibit peak ratios below unity. Black hole feedback might have already quenched these galaxies, thereby removing their ability to add new stars. Stellar mass loss can then occur either due to stellar evolution or tidal stripping. Furthermore, there is a clear shift with host mass: surviving satellites in more massive hosts are prone to lose parts of their stellar mass more easily. In cluster environments of $10^{14} - 10^{14.6}~\rmn{M}_\odot$ roughly 40 to 50 per cent of satellites show a net mass loss in their stellar mass components. However, since we only consider surviving satellites, those in less massive hosts that lost a larger fraction of their stellar mass since infall might simply have been disrupted. Satellites in more massive hosts, on the other hand, can be more massive themselves and can therefore lose a larger fraction of their stellar mass without falling beneath sample or resolution limits. Similar trends also hold for the alternative sample of surviving satellites that were selected using their peak dynamical mass.

This picture is consistent with results from literature: \cite{smith2013} study the onset of stellar stripping. Using simulations of galaxies interacting with the gravitational potential of a Virgo-like cluster, they examine the remains of dark matter subhaloes at the point when 10 per cent of the satellites' stellar mass has been stripped. Comparing various galaxy models, the loss of stellar mass set in only after 15 to 20 per cent of the bound dark matter fraction was left.

\cite{smith2016} follow these results up by investigating tidal stripping of dark matter and stellar mass of low-mass satellites in high-resolution cosmological hydrodynamical simulations. While losing 70 per cent of dark matter to interactions with the cluster potential, the stellar component remains unaffected. By the time the satellite has been stripped of 84 per cent of its dark matter, only 10 per cent of its stellar mass has been removed. This results due to the larger extent of dark matter subhaloes (compared to the galaxy itself). Comparing stellar-to-halo size-ratios and mass loss for extended and concentrated galaxies, both \cite{smith2016} and \cite{chang2013} find concentrated galaxies to be less likely to be stripped by their environment. In these galaxies, the stellar mass resides deeper inside the subhalo, so a larger fraction of dark matter has to be removed for it to be affected. While \cite{smith2016} find more massive galaxies to be more concentrated than low-mass galaxies -- and should therefore be able to retain more of their stellar mass --, galaxies in Figure~\ref{fig:massratios} exhibit the opposite trend. Massive satellite galaxies in TNG100 are actually more likely to be stripped of their stellar component than low-mass satellites.

Furthermore, \cite{bahe2019} find similar trends considering the mass loss of galaxies. They studied the survival and disruption of satellite galaxies in groups and clusters using cosmological zoom-in simulations and find stellar mass to be stripped to a lesser degree than total subhalo mass. Satellites tend to either retain a significant fraction of their stellar mass or are disrupted completely (i.e. quickly).

\subsection{Satellite SHMR shift as a function of host mass \& infall times}
\label{sec:infalltimes}

In Figure~\ref{fig:massratios}, it does not appear as if there is a significant variation in the strength of tidal stripping with host mass: therefore, the cause for the shift in the SHMR in Figure~\ref{fig:shmr_main2} remains to be determined. If the distribution of satellite infall times changes with host mass, the dependence of the satellite SHMR shift with host mass may simply reflect an effect of different typical infall times. We examine this in the following section.

We present the infall distributions of TNG300 satellites -- that survive to $z=0$ with at least $10^{10.5}~\rmn{M}_\odot$ in dynamical mass and that are found at $z=0$ within the virial radius of their host -- in three bins of host mass ($10^{12} - 10^{13}~\rmn{M}_\odot$, $10^{13} - 10^{14}~\rmn{M}_\odot$, $10^{14} - 10^{15.2}~\rmn{M}_\odot$; orange to red, solid curves) in the left panel of Figure~\ref{fig:hostmass_infall}. The infall distributions are smoothed using a Gaussian kernel with an average width of $0.3~\rmn{Gyr}$. Interestingly, the distribution of accretion times of surviving satellites is bimodal. This apparent bimodality of infall histories arises due to backsplash galaxies \citep{yun2019}. After first pericentric passage, the orbits of satellites can still extend outside their host's virial radius. However, since we define satellites to be within the virial radius, these galaxies are not part of our sample while they would otherwise fill up the infall time distributions at intermediate times (dotted curves). Regardless, the accretion of satellites peaks over the last $2.5~\rmn{Gyr}$ with a smaller, secondary peak 5--7 Gyr ago. This secondary peak is shifted to earlier times for satellites in more massive hosts, however, it is less pronounced for satellites in group-like hosts of $10^{12} - 10^{13}~\rmn{M}_\odot$. The infall times of satellites that survive through $z=0$ and now reside in lower-mass hosts span an overall smaller range of time, which could be a reason why these satellite populations exhibit on average smaller deviations from the centrals' SHMR. Including satellites outside the virial radius would not change our results nor the trends with host mass for the SHMR or its scatter. In fact, they would reinforce the trends with host mass in the left panel of Figure~\ref{fig:shmr_main2} by expanding the SHMR shifts more significantly for satellites with larger dynamical mass.

The trends found above also hold when we consider an alternative sample, i.e. selecting satellites that survive through $z=0$ by their peak instead of their present-day dynamical mass, as previously done in Section~\ref{sec:massloss} and Figure~\ref{fig:massratios}. In this case, most satellites fall into their present-day host environment's progenitor earlier in time, with a broader early infall time peak ranging between lookback times of $6-10~\rmn{Gyr}$. Most early infallers in this alternative sample experience a strong degree of tidal stripping, which brings them below the dynamical mass limit imposed at present time for our fiducial satellite sample. However, the trends with host mass are still the same, with satellites in lower-mass hosts exhibiting later infall times. On the other hand, if we were to inspect the infall time distributions of all satellites ever accreted -- so including not only the present-day, surviving satellite galaxies but all satellites with a peak dynamical mass of $M_\rmn{dyn, peak} \geq 10^{10.5}~\rmn{M}_\odot$ that have ever been accreted -- the infall times would appear somewhat differently. The infall distributions would cover the same range in time regardless of host mass, with low-mass hosts in fact peaking slightly earlier, rather than later, than more massive ones, consistent with the trends of halo formation time with halo mass. A significant fraction of satellites that fell in present-day groups and clusters early on, $8-12~\rmn{Gyr}$ ago, have been disrupted in the meantime. Therefore, the infall time distribution of surviving satellites in Figure~\ref{fig:hostmass_infall} is biased towards more recent cosmic epochs.

While there is a shift in the distribution of surviving satellite infall times with host mass, we still need to confirm whether this causes a shift in stellar mass fractions with host mass as in the left panel of Figure~\ref{fig:shmr_main2}. Therefore, we further examine the combined dependence on host mass and infall times in the right panel of Figure~\ref{fig:hostmass_infall}. This panel depicts the ratio of stellar mass fractions $M_*/M_\rmn{dyn}$ of satellites and centrals as a function of host mass in different bins of infall lookback time (0--1, 1--2, 2--4, 4--6, 6--8, 8--10, 10--12 Gyr ago; green to blue curves). Generally, even at fixed infall time, satellites exhibit an increasing offset from the SHMR of centrals with increasing host mass -- more massive clusters are in fact more efficient in driving satellites to larger stellar mass fractions. However, there is also a clear trend with infall time: the earliest infallers (10--12 Gyr ago) in the most massive hosts can reach stellar mass fractions of up to a factor 100 larger than those of centrals. On the other hand, satellites in the most recent infall time bins (0--1 and 1--2 Gyr ago) exhibit significantly lower ratios of stellar mass fractions than satellites of all other infall times. These galaxies have not yet spent enough time inside their new host environment to have experienced extended stripping or even a pericentric passage.

\begin{figure*}
    \centering
    \includegraphics[width=.7\textwidth]{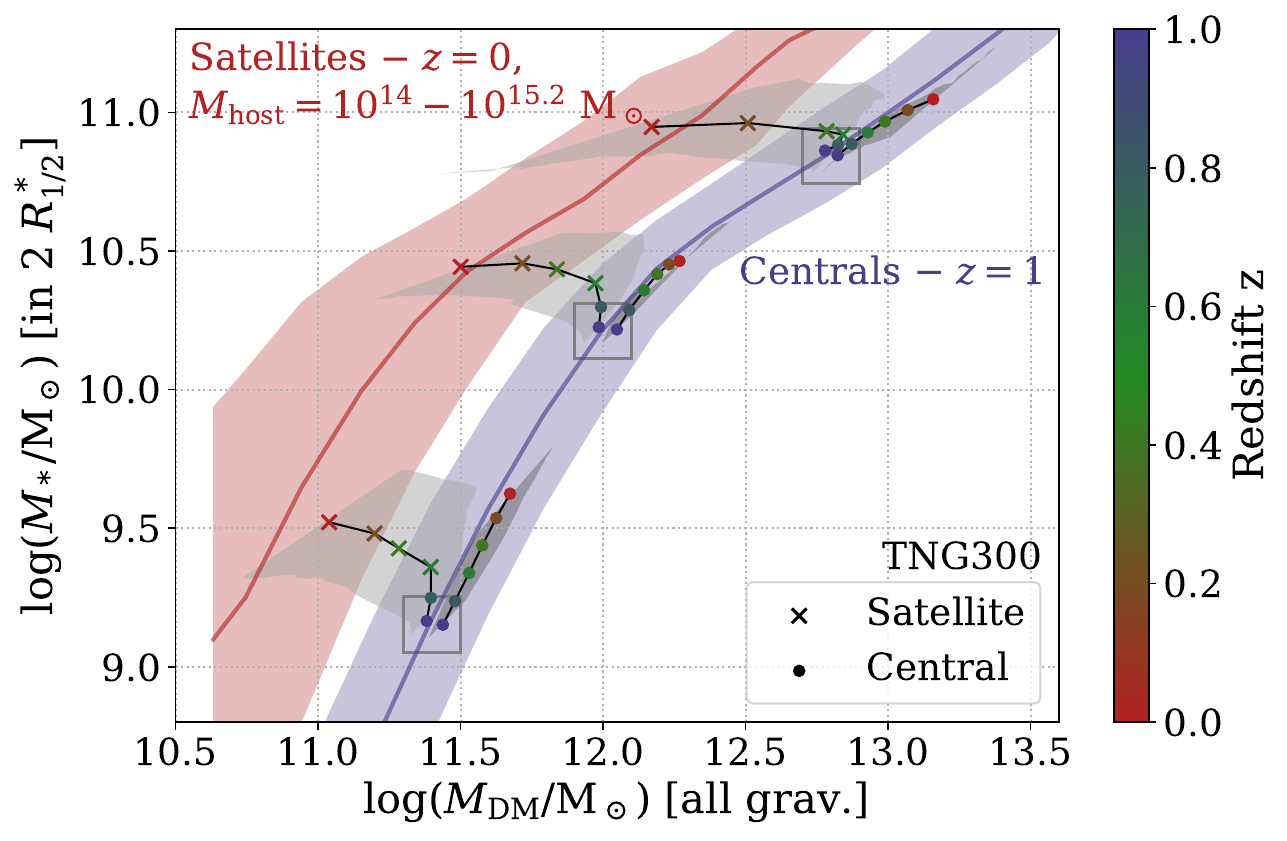}
    \caption{Evolution of stellar and dark matter mass in our fiducial aperture choice (all gravitationally bound dark matter particles) from $z=1$ to $z=0$ in TNG300. The blue curve corresponds to the SHMR for centrals at $z=1$, the red curve to the SHMR of satellites in massive cluster hosts of $10^{14} - 10^{15.2}~\rmn{M}_\odot$ at $z=0$. Shaded regions depict the scatter as $16^\rmn{th}$ and $84^\rmn{th}$ percentiles. We choose centrals at $z=1$ within a parameter space of $M_\rmn{DM} = 10^{11.3} - 10^{11.5}~\rmn{M}_\odot$ and $M_* = 10^{9.05} - 10^{9.25}~\rmn{M}_\odot$, $M_\rmn{DM} = 10^{11.9} - 10^{12.1}~\rmn{M}_\odot$ and $M_* = 10^{10.11} - 10^{10.31}~\rmn{M}_\odot$, or $M_\rmn{DM} = 10^{12.7} - 10^{12.9}~\rmn{M}_\odot$ and $M_* = 10^{10.74} - 10^{10.94}~\rmn{M}_\odot$ (denoted by grey boxes), and follow their evolution to $z=0$. Median evolutionary tracks are shown separately for galaxies that stay centrals or become satellites: centrals are denoted by dots, satellites by crosses. The markers are colour-coded by time, covering redshift $z = 1.0, 0.8, 0.6, 0.4, 0.2, 0.0$. Here, infall is defined as the first time the subhaloes are considered members of another FoF halo (\textit{not} infall into its virial radius $R_\rmn{200c}$ as for the rest of this paper). Grey shaded areas correspond to the scatter of central and satellite evolutionary tracks as $16^\rmn{th}$ and $84^\rmn{th}$ percentiles.}
    \label{fig:evo_tracks}
\end{figure*}
\subsection{Evolution of centrals and satellites in the stellar mass vs. halo mass plane}
\label{sec:evotracks}

In order to illustrate the differences in the evolution of centrals and satellites, as well as the contributions of ongoing star formation and tidal stripping in the host potential, we present the SHMR of TNG300 as stellar mass versus dark matter mass and the progression of galaxies between $z=1$ and $z=0$ in Figure~\ref{fig:evo_tracks}. Here, we consider dark matter instead of dynamical mass in order to illustrate the impact of tidal stripping on haloes directly. While gas stripping does occur -- especially for dwarf galaxies with larger gas fractions -- it is negligible compared to the loss of dark matter. As seen in Figure~\ref{fig:massratios}, the distribution of dark matter mass loss traces the distribution of dynamical mass.

We consider the SHMR of centrals at $z=1$ (blue curve) and compare it to the SHMR of satellites in massive clusters of $10^{14} - 10^{15.2}~\rmn{M_\odot}$ at $z=0$. At $z=1$, we define various parameter spaces in the SHMR (denoted by the grey boxes) and select two disjoint sets of galaxies in each bin -- depending on whether they stay centrals or become satellites by $z=0$. Their average evolutionary tracks are depicted at $z=1.0, 0.8, 0.6, 0.4, 0.2, 0.0$ using median stellar and dark matter mass at the respective points in time. Markers show whether the galaxies are centrals (dots) or have become a satellite as member of another FoF halo (crosses). 

Centrals remain undisturbed by the environment, grow more massive in both stellar and dark matter mass, and evolve more or less along the same $z=1$ SHMR. The evolutionary tracks of satellites, however, present a different picture: in the low- and intermediate-mass bins, their dark matter growth is reduced and halted even while they are still considered centrals. Their relatively nearby, future host halo possibly already dominates the accretion of dark matter since mass accretion for clusters persists out to several virial radii \citep{behroozi2014}. Star formation continues and they begin to move off their original SHMR in an almost vertical fashion.

In the massive bin, galaxies still evolve along the SHMR during this first phase: their star formation may be already quenched, in our model via AGN feedback \citep[e.g.][]{weinberger2017, donnari2019, donnari2020a, terrazas2020}, and they primarily grow due to mergers with other galaxies. However, as soon as galaxies become satellites of a more massive halo, tidal stripping by the potential of the new host removes the outer parts of the satellite galaxies' dark matter subhaloes and dominates the transition to the SHMR of satellites until $z=0$ -- irrespective of their dynamical or stellar mass. The star formation activity of galaxies in the low- and intermediate-mass bins decreases after infall. While the scatter for the evolutionary tracks (grey shaded areas) is fairly broad with up to $\sim0.3-0.4~\rm{dex}$ at fixed dynamical mass, the tracks of the $16^\rmn{th}$ and $84^\rmn{th}$ percentile populations follow the same trends -- shifted to lower or higher stellar masses, respectively. This scatter might be introduced by different orbital configurations or initial pericentric distances. However, we do not find significant stripping of stellar mass in the average satellite evolution tracks (as already evident from Figure~\ref{fig:massratios}).

\cite{niemiec2019} found similar results in the Illustris simulation: after infall, satellite galaxies in massive clusters can be stripped of up to 80 per cent of their dark matter subhalo after spending 8--9 Gyr in their host. Furthermore, these satellites continue to form stars until they experience their first pericentric passage. They interpret the shift in the SHMR of satellite galaxies to result from three different phases: (i) loss of dark matter by tidal stripping and increase in stellar mass by star formation, (ii) loss of dark matter and constant stellar mass after quenching, as well as (iii) combined loss of dark matter and stellar mass by tidal stripping. While we recover trends similar to the first two phases for the transition of satellite galaxies, we do not find a significant combined loss of dark matter and stellar mass for TNG galaxies. These differences might arise due to different galaxy formation models: low-mass galaxies in Illustris have been found to be too large by a factor of $\sim 2-3$ in comparison to observations \citep{snyder2015} and IllustrisTNG \citep{pillepich2018a}. Due to their increased extent, the stellar component of these galaxies may become subject to tidal stripping more easily.

\subsection{Tools and fitting functions}
\label{sec:tools}

We provide a family of fitting functions for the SHMR in IllustrisTNG. As for Figure~\ref{fig:shmr_main2}, these functions are constructed using the combined sample of rTNG300, rTNG100, and TNG50. We adopt a parametrization similar to \cite{moster2010,moster2013} as per Equation~\eqref{eq:mosterfit}. We summarise the parameters for the best fitting models for dynamical and stellar masses in our fiducial aperture choice (all gravitationally bound particles for $M_\rmn{dyn}$, stellar mass within two stellar half-mass radii for $M_*$) in Table~\ref{tab:disc_fits_fidAp}. Since satellites in different environments form different SHMRs, Table~\ref{tab:disc_fits_fidAp} includes variations of host mass range and bin sizes. In Appendix~\ref{sec:app_fits}, we visualise how the fitting parameters vary with host halo mass. 

\begin{table*}
    \centering
    \begin{tabular}{l c c c c c}
        $\mathbf{M_\rmn{dyn}}$ \textbf{[all grav.],} $\mathbf{M_\rmn{*}}$ \textbf{[in 2 $\mathbf{R_{1/2}^*}$]}\\
        \hline \hline
        Sample & Median $M_\rmn{host}~[\log M_\odot]$ & $N$ & $M_1~[\log \rmn{M}_\odot]$ & $\beta$ & $\gamma$\\ \hline
        Centrals & & $0.0258 \pm 0.0003$ & $11.70 \pm 0.02$ & $28.6 \pm 0.8$ & $10.4 \pm 0.2$\\
        Satellites in $10^{12} - 10^{13}~\rmn{M}_\odot$ hosts & 12.53 & $0.108 \pm 0.003$ & $11.12 \pm 0.06$ & $27.5 \pm 2.9$ & $15.6 \pm 1.7$\\
        Satellites in $10^{13} - 10^{14}~\rmn{M}_\odot$ hosts & 13.52 & $0.127 \pm 0.008$ & $10.85 \pm 0.11$ & $23.6 \pm 6.1$ & $10.1 \pm 1.3$\\
        Satellites in $10^{14} - 10^{14.5}~\rmn{M}_\odot$ hosts & 14.22 & $0.137 \pm 0.004$ & $10.93 \pm 0.04$ & $30.5 \pm 3.6$ & $10.9 \pm 0.6$\\
        Satellites in $10^{14.5} - 10^{15.2}~\rmn{M}_\odot$ hosts & 14.67 & $0.129 \pm 0.006$ & $10.85 \pm 0.04$ & $38.6 \pm 7.2$ & $9.5 \pm 0.7$\\\hline\\
        Satellites in $10^{12} - 10^{12.5}~\rmn{M}_\odot$ hosts & 12.27 & $0.092 \pm 0.010$ & $11.17 \pm 0.27$ & $25.4 \pm 8.8$ & $16.4 \pm 10.3$\\
        Satellites in $10^{12.5} - 10^{13}~\rmn{M}_\odot$ hosts & 12.75 & $0.123 \pm 0.004$ & $11.00 \pm 0.05$ & $33.6 \pm 3.5$ & $15.3 \pm 1.4$\\
        Satellites in $10^{13} - 10^{13.5}~\rmn{M}_\odot$ hosts & 13.26 & $0.130 \pm 0.003$ & $11.12 \pm 0.22$ & $16.5 \pm 5.8$ & $13.2 \pm 2.6$\\
        Satellites in $10^{13.5} - 10^{14}~\rmn{M}_\odot$ hosts & 13.76 & $0.128 \pm 0.009$ & $10.76 \pm 0.08$ & $28.5 \pm 7.8$ & $9.3 \pm 1.1$\\\hline\\
        Satellites in $10^{12.5} - 10^{13.5}~\rmn{M}_\odot$ hosts & 13.01 & $0.114 \pm 0.005$ & $10.88 \pm 0.06$ & $33.3 \pm 5.4$ & $10.5 \pm 1.0$\\
        Satellites in $10^{13.5} - 10^{14.5}~\rmn{M}_\odot$ hosts & 13.94 & $0.133 \pm 0.005$ & $10.86 \pm 0.05$ & $28.00 \pm 4.3$ & $10.4 \pm 0.7$\\
        \hline\\
    \end{tabular}
    \caption[]{Fit parameters to the SHMR of centrals and satellites in various bins of host mass using rTNG300, rTNG100 and TNG50 galaxies and our fiducial aperture choice (all gravitationally bound mass for $M_\rmn{dyn}$, stellar mass within twice the stellar half-mass radius for $M_*$). We follow the parametrization in Equation~\eqref{eq:mosterfit}, using normalisation $N$, characteristic mass $M_1$, and the slopes at the low- and high-mass ends $\beta$ and $\gamma$.}
    \label{tab:disc_fits_fidAp}
\end{table*}

\subsection{Halo finder and resolution limitations}
\label{sec:halofinder}

The results uncovered so far represent the outcome of the numerical galaxy formation model as implemented in IllustrisTNG and it may be that other cosmological simulations in the future will return somewhat different quantitative (albeit -- we believe -- not qualitative) solutions. In practice, also within the IllustrisTNG simulations, our quantitative results may depend to some extent on the underlying adopted identification tools as well as on the underlying numerical resolution. 

In what follows, we want to discuss the limitations and possible tensions for the measurement of dynamical masses accomplished thanks to the \textsc{subfind} algorithm \citep{springel2001, dolag2009}. By using all gravitationally bound particles for the subhalo masses, we rely on the way resolution elements (or particles) are assigned by the halo finder to subhaloes and there may be physical situations whereby such assignment can be difficult or problematic. It should be noticed from the onset that, although \textsc{subfind} defines a subhalo as the collection of a certain minimum number of particles that survive the unbinding procedure, the choice of 20 as minimum number of resolution elements per subhalo adopted here cannot constitute an issue, as throughout the analysis we only consider galaxies with minimum dynamical masses of $10^{10.5}~\rmn{M}_\odot$ (i.e. at least many hundreds of particles for satellites even at the lowest resolution adopted in this paper). 

\textsc{subfind} identifies substructure within a parent FoF halo as groups of particles that form gravitationally self-bound, locally overdense regions. Subhaloes in locations of generally higher density -- such as areas close to the centres of host haloes -- could be misidentified or have underestimated dynamical masses, with parts of their outskirts being ascribed to their centrals. We avoid these regions by imposing a minimum cluster-centric distance on our satellite sample: only satellites that are located at least $0.05~R_\rmn{200c}$ from their host's centre are included. However, we have verified that not imposing this minimum cluster-centric distance does not change our results significantly.

Close objects might also lead to discrepancies. If two galaxies are situated too near to one another -- e.g. in a fly-by event -- the algorithm might run into problems separating them, since it only probes for local overdensities. However, considering the statistical size of our samples, we do not expect this to affect our findings.

\cite{ayromlou2019} constructed an instantaneous technique to identify additional member particles of subhaloes in their local background environment. Using a Gaussian mixture method, they classify background particles into two components depending on whether particles share mean velocities and velocity dispersions similar to the original subhalo. These particles are then reassigned to the subhaloes in order to decontaminate the true background particles. This results in a noticeable effect on the satellite stellar mass function: masses of subhaloes can increase by factors of 2 or more. Mass changes are larger for more massive satellites and -- at fixed subhalo mass -- larger for satellites in lower-mass hosts.

Generally, these possible uncertainties could be alleviated at once by employing and comparing to another halo finder. 6D halo finders such as \textsc{rockstar} \citep{behroozi2013} or \textsc{VELOCIraptor} \citep{elahi2019} additionally take velocity information into account to identify substructures. This might yield different dynamical masses for this study, however, we do not expect this to change the qualitative trends in our results. While comparing the identification of environmental effects and tidal stripping of satellite galaxies between different halo finders might yield additional insights, it exceeds the scope of this study.

While our galaxy sample seems relatively safe regarding limitations in the identification of halo overdensities and substructures, satellites might become subject to artificial disruption because of the limited numerical resolution. When comparing our results across all the resolution levels of the IllustrisTNG suite, we find some dependence on numerical resolution, which is the reason why we present our results after applying a resolution correction that is gauged to reproduce quantitative results coherent with those from our highest-resolution realization: TNG50 (see Appendix~\ref{sec:resc}). 

However, by studying the evolution of satellite dark matter subhaloes in a series of idealized N-body simulations, \cite{vandenbosch2018} found most tidal disruption events to be of numerical origin and that inadequate force softening (as that adopted in typical cosmological large-volume simulations like TNG100 or TNG300) can lead to overestimated mass loss. However, a number of caveats makes it difficult to extrapolate these findings to more realistic cosmological setups: those results are based on dark matter-only simulations (i.e. without contributions of baryonic effects), satellites are bound to purely circular, infinitely-long orbits, dynamical friction is not accounted for, and the host halo is represented by a static analytical potential. In fact, \cite{bahe2019} relax some of these concerns by studying the survival rate of satellite galaxies in cosmological zoom-in simulations. According to their findings, total disruption of satellites is negligible in massive clusters and predominantly occurs in lower-mass groups and during preprocessing. Furthermore, the disruption efficiency shows a strong correlation with redshift: the fraction of surviving satellites decreases towards earlier accretion times and is in any case physically negligible for accretion times of $z \gtrsim 4$. This is consistent with our findings in Figure~\ref{fig:hostmass_infall}. Furthermore, \cite{bahe2019} find that while baryons contribute to the degree of mass loss satellite galaxies experience, they only have a small impact on their actual rate of survival. Whether subhaloes are artificially over-stripped or completely destroyed might correspond to different physical problems. While \cite{vandenbosch2018} focus on the possibly artificial, {\it complete} disruption of subhaloes (i.e. overmerging), their results considering the actual amount of mass stripped are reassuring within the context of ``low-resolution'' cosmological simulations. According to their figure 10, the first 99 per cent of material stripped from a subhalo is perfectly well captured -- also at the resolutions that are relevant here. 

\section{Summary and conclusions}
\label{sec:conc}

We have analysed the stellar-to-halo mass relation (SHMR) in the suite of cosmological magneto-hydrodynamical simulations IllustrisTNG, using all three flagship runs TNG50, TNG100, and TNG300. We distinguished between centrals and satellites with total dynamical masses of $M_\rmn{dyn} \geq 10^{10.5}~\rmn{M}_\odot$ and considered exclusively satellites in group- and cluster-like hosts with $M_\rmn{host} = 10^{12} - 10^{15.2}~\rmn{M}_\odot$. We have characterised the effects of such environments on the evolution of galaxies, their surrounding dark matter subhaloes, and the SHMR scatter as a function of total dynamical mass. We have combined the results of all three IllustrisTNG simulations to maximise the dynamic range and have devised a resolution correction of the galaxy stellar masses that extrapolates the TNG100 and TNG300 results to TNG50 resolution, resulting in three sets of output with the same effective numerical mass resolution. Our results are summarised as follows.

\begin{itemize}
    \item The SHMR of satellite galaxies in groups and clusters of at least $M_\rmn{host} \geq 10^{12}~\rmn{M}_\odot$ exhibits a significant offset from the SHMR of centrals (Figure~\ref{fig:shmr_main}). At fixed $z=0$ dynamical mass, satellites have larger stellar masses and larger stellar mass fractions. This shift and the scatter of the relation correlates with the mass of their host: for example, satellites in hosts of $10^{14} - 10^{15.2}~\rmn{M}_\odot$ at $z=0$ reach median stellar mass fractions of up to 15 per cent at the SHMR's peak, while satellites in less massive hosts of $10^{12} - 10^{13}~\rmn{M}_\odot$ reach only 10 per cent (Figure~\ref{fig:shmr_main2}, left panel). This is a significant difference compared to centrals, which display a peak stellar mass fraction of about 2--4 per cent.\\
    \item This offset between the SHMRs of central and satellite galaxies is the result of environmental effects that act in an outside-in fashion. Since the inner galaxy regions remain largely unaffected by their environment, the offset between the SHMRs of centrals and satellites disappears if we measure masses within sufficiently small physical apertures (Figure~\ref{fig:shmr_main}, bottom panels).\\
    \item The ratio of stellar mass between satellites and centrals as a function of total dynamical mass for satellites within their host's virial radius $R_\rmn{200c}$ increases towards lower dynamical mass (up to a factor of 16 at $z=0$) and shows no significant evolution with time in the range $z=0-2$ (Figure~\ref{fig:shmr_main2}, right panel). The tidal forces within the host halo's gravitational potential strip a significant fraction of satellite subhaloes over relatively short time scales.\\ 
    \item While the scatter $\sigma_*$ in (logarithmic) stellar mass as a function of dynamical mass of both centrals and satellites follows the same shape -- roughly constant at $0.1-0.2~\rmn{dex}$ for dynamical masses above the respective SHMR peak, and increasing towards the lower mass end -- satellites exhibit a higher scatter over the whole range of dynamical mass (Figure~\ref{fig:mstar_scatter_TNG100-300}). However, the rise in scatter at low subhalo masses is steeper for satellites than for centrals since these dwarf-like satellites are more susceptible to the impact of group and cluster environments. Here, $\sigma_*$ reaches up to $0.6-0.8~\rmn{dex}$ for the least massive galaxies considered. The SHMR scatter of the mass-limited sample of satellites increases continuously with increasing host mass. Satellites with $M_\rmn{dyn} \gtrsim 10^{12}~\rmn{M}_\odot$ show no evolution with redshift. For satellites of lower dynamical mass, however, the scatter decreases systematically with increasing redshift -- albeit only weakly (Figure~\ref{fig:mstar_scatter_TNG100-300}, bottom right panel).\\
    \item At fixed $z=0$ dynamical masses, satellites with higher apparent stellar mass fractions tend to reside closer to the group or cluster centre, experienced an earlier infall (both into the virial radius of their present-day host and into another halo in general), and inhabit higher local luminosity density regions than analog satellites with lower stellar mass fractions (Figures~\ref{fig:shmr_env_allsats} and \ref{fig:shmr_env_offset}). \\
    \item Infall into a more massive environment exerts distinct impacts on the dark matter and stellar components of satellite galaxies (Figure~\ref{fig:massratios}). While dark matter mass is dominated by tidal stripping and overall mass loss -- regardless of host mass or the satellites' stellar mass -- there is a significant net increase for stellar mass and still ongoing star formation post-infall. However, the stellar mass distribution shifts towards net mass loss with both increasing host mass and galaxy stellar mass. Tidal stripping of stars becomes more efficient within the deeper potentials of massive galaxy clusters. Since more massive galaxies might already be quenched pre-infall, they show a less distinct net mass gain.\\
    \item More massive clusters are more efficient in driving satellites to larger stellar mass fractions (Figure~\ref{fig:hostmass_infall}). Satellites that survive through $z=0$ in lower-mass hosts cover a smaller range of infall times compared to satellite populations in more massive hosts -- and are therefore exposed to their host environment for a shorter time. Furthermore, as noted above, satellites with earlier infall time have been exposed to the cluster/group potential for a longer time and generally exhibit larger SHMR offsets from central galaxies. Yet, even at fixed infall time, the stellar mass fractions of satellites exhibit an increasing offset with host mass compared to the SHMR of centrals.\\
    \item Considering the evolution of centrals into satellites in the SHMR plane between $z=1$ and $z=0$ (Figure~\ref{fig:evo_tracks}), we find the transition to be dominated by dark matter loss and tidal stripping after star formation has been quenched by the infall into a more massive host. However, even before the galaxies have become satellites they start to move off the centrals' SHMR due to a decreasing growth in dark matter and continued star formation. Galaxies that stay centrals, on the other hand, simply evolve along the SHMR (which evolves only weakly at $z<1$) and increase in both stellar and dark matter mass.
\end{itemize}

In conclusion, we have highlighted the influence of group and cluster environments on the stellar and dynamical mass components of satellite galaxies. Satellite galaxies selected at a given time with a certain minimum dynamical or total mass do not simply contribute to the scatter in the SHMR of central galaxies but form their own distinct, separate relation. Whether they become satellites of a low-mass group or of a massive galaxy cluster, their SHMR shifts and their scatter increases with respect to the SHMR of centrals. While satellites might appear to be more efficient at forming stars when compared to centrals at fixed total dynamical mass, this difference is predominantly caused by tidal stripping of their dark subhaloes by the gravitational potential of a more massive host halo.

\section*{Data Availability}
The TNG300 and TNG100 simulations of IllustrisTNG are publicly available at \url{www.tng-project.org/data}; TNG50 will become public in the future. Data directly referring to content and figures of this publication is available upon request from the corresponding author.

\section*{Acknowledgements}
CE acknowledges support by the Deutsche Forschungsgemeinschaft (DFG, German Research Foundation) through project 394551440 and thanks Elad Zinger for sharing catalogs of backsplash galaxies in IllustrisTNG, which were used for additional comparisons to address points in the referee report.
FM acknowledges support through the Program "Rita Levi Montalcini" of the Italian MIUR.
The flagship simulations of the IllustrisTNG project used in this work have been run on the HazelHen Cray XC40-system at the High Performance Computing Center Stuttgart as part of project GCS-ILLU (PI: Springel) and GCS-DWAR (Co-PIs: Nelson, Pillepich) of the Gauss centres for Super-computing (GCS). Ancillary and test runs of the project were also run on the Stampede supercomputer at TACC/XSEDE (allocation AST140063), at the Hydra and Draco supercomputers at the Max Planck Computing and Data Facility, and on the MIT/Harvard computing facilities supported by FAS and MIT MKI.



\bibliographystyle{mnras}
\bibliography{engler2020_shmr}

\begin{thebibliography}{}
\makeatletter
\relax
\def\mn@urlcharsother{\let\do\@makeother \do\$\do\&\do\#\do\^\do\_\do\%\do\~}
\def\mn@doi{\begingroup\mn@urlcharsother \@ifnextchar [ {\mn@doi@}
  {\mn@doi@[]}}
\def\mn@doi@[#1]#2{\def\@tempa{#1}\ifx\@tempa\@empty \href
  {http://dx.doi.org/#2} {doi:#2}\else \href {http://dx.doi.org/#2} {#1}\fi
  \endgroup}
\def\mn@eprint#1#2{\mn@eprint@#1:#2::\@nil}
\def\mn@eprint@arXiv#1{\href {http://arxiv.org/abs/#1} {{\tt arXiv:#1}}}
\def\mn@eprint@dblp#1{\href {http://dblp.uni-trier.de/rec/bibtex/#1.xml}
  {dblp:#1}}
\def\mn@eprint@#1:#2:#3:#4\@nil{\def\@tempa {#1}\def\@tempb {#2}\def\@tempc
  {#3}\ifx \@tempc \@empty \let \@tempc \@tempb \let \@tempb \@tempa \fi \ifx
  \@tempb \@empty \def\@tempb {arXiv}\fi \@ifundefined
  {mn@eprint@\@tempb}{\@tempb:\@tempc}{\expandafter \expandafter \csname
  mn@eprint@\@tempb\endcsname \expandafter{\@tempc}}}

\bibitem[\protect\citeauthoryear{Allen, Behroozi  \& Ma}{Allen
  et~al.}{2019}]{allen2019}
Allen M.,  Behroozi P.,   Ma C.-P.,  2019, \mn@doi [\mnras]
  {10.1093/mnras/stz2067}, 488, 4916

\bibitem[\protect\citeauthoryear{{Anbajagane}, {Evrard}, {Farahi}, {Barnes},
  {Dolag}, {McCarthy}, {Nelson}  \& {Pillepich}}{{Anbajagane}
  et~al.}{2020}]{anbajagane2020}
{Anbajagane} D.,  {Evrard} A.~E.,  {Farahi} A.,  {Barnes} D.~J.,  {Dolag} K.,
  {McCarthy} I.~G.,  {Nelson} D.,   {Pillepich} A.,  2020, arXiv e-prints,
  \href {https://ui.adsabs.harvard.edu/abs/2020arXiv200102283A} {p.
  arXiv:2001.02283}

\bibitem[\protect\citeauthoryear{{Ashman}, {Salucci}  \& {Persic}}{{Ashman}
  et~al.}{1993}]{ashman1993}
{Ashman} K.~M.,  {Salucci} P.,   {Persic} M.,  1993, \mn@doi [\mnras]
  {10.1093/mnras/260.3.610}, \href
  {https://ui.adsabs.harvard.edu/abs/1993MNRAS.260..610A} {260, 610}

\bibitem[\protect\citeauthoryear{{Ayromlou}, {Nelson}, {Yates}, {Kauffmann}  \&
  {White}}{{Ayromlou} et~al.}{2019}]{ayromlou2019}
{Ayromlou} M.,  {Nelson} D.,  {Yates} R.~M.,  {Kauffmann} G.,   {White} S.
  D.~M.,  2019, \mn@doi [\mnras] {10.1093/mnras/stz1549}, \href
  {https://ui.adsabs.harvard.edu/abs/2019MNRAS.487.4313A} {487, 4313}

\bibitem[\protect\citeauthoryear{{Bah{\'e}}, {McCarthy}, {Balogh}  \&
  {Font}}{{Bah{\'e}} et~al.}{2013}]{bahe2013}
{Bah{\'e}} Y.~M.,  {McCarthy} I.~G.,  {Balogh} M.~L.,   {Font} A.~S.,  2013,
  \mn@doi [\mnras] {10.1093/mnras/stt109}, \href
  {https://ui.adsabs.harvard.edu/\#abs/2013MNRAS.430.3017B} {430, 3017}

\bibitem[\protect\citeauthoryear{{Bah{\'e}} et~al.,}{{Bah{\'e}}
  et~al.}{2017}]{bahe2017}
{Bah{\'e}} Y.~M.,  et~al., 2017, \mn@doi [\mnras] {10.1093/mnras/stx1403},
  \href {https://ui.adsabs.harvard.edu/abs/2017MNRAS.470.4186B} {470, 4186}

\bibitem[\protect\citeauthoryear{{Bah{\'e}} et~al.,}{{Bah{\'e}}
  et~al.}{2019}]{bahe2019}
{Bah{\'e}} Y.~M.,  et~al., 2019, \mn@doi [\mnras] {10.1093/mnras/stz361}, \href
  {https://ui.adsabs.harvard.edu/abs/2019MNRAS.485.2287B} {485, 2287}

\bibitem[\protect\citeauthoryear{{Balogh}, {Morris}, {Yee}, {Carlberg}  \&
  {Ellingson}}{{Balogh} et~al.}{1999}]{balogh1999}
{Balogh} M.~L.,  {Morris} S.~L.,  {Yee} H.~K.~C.,  {Carlberg} R.~G.,
  {Ellingson} E.,  1999, \mn@doi [\apj] {10.1086/308056}, \href
  {https://ui.adsabs.harvard.edu/abs/1999ApJ...527...54B} {527, 54}

\bibitem[\protect\citeauthoryear{{Balogh}, {Navarro}  \& {Morris}}{{Balogh}
  et~al.}{2000}]{balogh2000}
{Balogh} M.~L.,  {Navarro} J.~F.,   {Morris} S.~L.,  2000, \mn@doi [\apj]
  {10.1086/309323}, \href
  {https://ui.adsabs.harvard.edu/abs/2000ApJ...540..113B} {540, 113}

\bibitem[\protect\citeauthoryear{{Barnes} \& {Hernquist}}{{Barnes} \&
  {Hernquist}}{1992}]{barnes1992}
{Barnes} J.~E.,  {Hernquist} L.,  1992, \mn@doi [\nat] {10.1038/360715a0},
  \href {https://ui.adsabs.harvard.edu/\#abs/1992Natur.360..715B} {360, 715}

\bibitem[\protect\citeauthoryear{{Behroozi}, {Conroy}  \&
  {Wechsler}}{{Behroozi} et~al.}{2010}]{behroozi2010}
{Behroozi} P.~S.,  {Conroy} C.,   {Wechsler} R.~H.,  2010, \mn@doi [\apj]
  {10.1088/0004-637X/717/1/379}, \href
  {https://ui.adsabs.harvard.edu/\#abs/2010ApJ...717..379B} {717, 379}

\bibitem[\protect\citeauthoryear{{Behroozi}, {Wechsler}  \& {Wu}}{{Behroozi}
  et~al.}{2013}]{behroozi2013}
{Behroozi} P.~S.,  {Wechsler} R.~H.,   {Wu} H.-Y.,  2013, \mn@doi [\apj]
  {10.1088/0004-637X/762/2/109}, \href
  {https://ui.adsabs.harvard.edu/abs/2013ApJ...762..109B} {762, 109}

\bibitem[\protect\citeauthoryear{{Behroozi}, {Wechsler}, {Lu}, {Hahn}, {Busha},
  {Klypin}  \& {Primack}}{{Behroozi} et~al.}{2014}]{behroozi2014}
{Behroozi} P.~S.,  {Wechsler} R.~H.,  {Lu} Y.,  {Hahn} O.,  {Busha} M.~T.,
  {Klypin} A.,   {Primack} J.~R.,  2014, \mn@doi [\apj]
  {10.1088/0004-637X/787/2/156}, \href
  {https://ui.adsabs.harvard.edu/\#abs/2014ApJ...787..156B} {787, 156}

\bibitem[\protect\citeauthoryear{{Bekki}}{{Bekki}}{2014}]{kenji2014}
{Bekki} K.,  2014, \mn@doi [\mnras] {10.1093/mnras/stt2216}, \href
  {https://ui.adsabs.harvard.edu/abs/2014MNRAS.438..444B} {438, 444}

\bibitem[\protect\citeauthoryear{{Binggeli}, {Tammann}  \&
  {Sandage}}{{Binggeli} et~al.}{1987}]{binggeli1987}
{Binggeli} B.,  {Tammann} G.~A.,   {Sandage} A.,  1987, \mn@doi [\aj]
  {10.1086/114467}, \href
  {https://ui.adsabs.harvard.edu/\#abs/1987AJ.....94..251B} {94, 251}

\bibitem[\protect\citeauthoryear{{Boselli} \& {Gavazzi}}{{Boselli} \&
  {Gavazzi}}{2006}]{boselli2006}
{Boselli} A.,  {Gavazzi} G.,  2006, \mn@doi [\pasp] {10.1086/500691}, \href
  {https://ui.adsabs.harvard.edu/abs/2006PASP..118..517B} {118, 517}

\bibitem[\protect\citeauthoryear{{Boylan-Kolchin}, {Springel}, {White},
  {Jenkins}  \& {Lemson}}{{Boylan-Kolchin} et~al.}{2009}]{boylan-kolchin2009}
{Boylan-Kolchin} M.,  {Springel} V.,  {White} S. D.~M.,  {Jenkins} A.,
  {Lemson} G.,  2009, \mn@doi [\mnras] {10.1111/j.1365-2966.2009.15191.x},
  \href {https://ui.adsabs.harvard.edu/\#abs/2009MNRAS.398.1150B} {398, 1150}

\bibitem[\protect\citeauthoryear{{Bradshaw}, {Leauthaud}, {Hearin}, {Huang}  \&
  {Behroozi}}{{Bradshaw} et~al.}{2020}]{bradshaw2020}
{Bradshaw} C.,  {Leauthaud} A.,  {Hearin} A.,  {Huang} S.,   {Behroozi} P.,
  2020, \mn@doi [\mnras] {10.1093/mnras/staa081}, \href
  {https://ui.adsabs.harvard.edu/abs/2020MNRAS.493..337B} {493, 337}

\bibitem[\protect\citeauthoryear{{Buck}, {Macci{\`o}}, {Dutton}, {Obreja}  \&
  {Frings}}{{Buck} et~al.}{2019}]{buck2019}
{Buck} T.,  {Macci{\`o}} A.~V.,  {Dutton} A.~A.,  {Obreja} A.,   {Frings} J.,
  2019, \mn@doi [\mnras] {10.1093/mnras/sty2913}, \href
  {https://ui.adsabs.harvard.edu/abs/2019MNRAS.483.1314B} {483, 1314}

\bibitem[\protect\citeauthoryear{{Chang}, {Macci{\`o}}  \& {Kang}}{{Chang}
  et~al.}{2013}]{chang2013}
{Chang} J.,  {Macci{\`o}} A.~V.,   {Kang} X.,  2013, \mn@doi [\mnras]
  {10.1093/mnras/stt434}, \href
  {https://ui.adsabs.harvard.edu/abs/2013MNRAS.431.3533C} {431, 3533}

\bibitem[\protect\citeauthoryear{{Cowie} \& {Songaila}}{{Cowie} \&
  {Songaila}}{1977}]{cowie1977}
{Cowie} L.~L.,  {Songaila} A.,  1977, \mn@doi [\nat] {10.1038/266501a0}, \href
  {https://ui.adsabs.harvard.edu/abs/1977Natur.266..501C} {266, 501}

\bibitem[\protect\citeauthoryear{{Dolag}, {Borgani}, {Murante}  \&
  {Springel}}{{Dolag} et~al.}{2009}]{dolag2009}
{Dolag} K.,  {Borgani} S.,  {Murante} G.,   {Springel} V.,  2009, \mn@doi
  [\mnras] {10.1111/j.1365-2966.2009.15034.x}, \href
  {https://ui.adsabs.harvard.edu/#abs/2009MNRAS.399..497D} {399, 497}

\bibitem[\protect\citeauthoryear{{Donnari} et~al.,}{{Donnari}
  et~al.}{2019}]{donnari2019}
{Donnari} M.,  et~al., 2019, \mn@doi [\mnras] {10.1093/mnras/stz712}, \href
  {http://adsabs.harvard.edu/abs/2019MNRAS.485.4817D} {485, 4817}

\bibitem[\protect\citeauthoryear{{Donnari} et~al.,}{{Donnari}
  et~al.}{2021a}]{donnari2020a}
{Donnari} M.,  et~al., 2021a, \mn@doi [\mnras] {10.1093/mnras/staa3006}, \href
  {https://ui.adsabs.harvard.edu/abs/2021MNRAS.500.4004D} {500, 4004}

\bibitem[\protect\citeauthoryear{{Donnari}, {Pillepich}, {Nelson}, {Marinacci},
  {Vogelsberger}  \& {Hernquist}}{{Donnari} et~al.}{2021b}]{donnari2020b}
{Donnari} M.,  {Pillepich} A.,  {Nelson} D.,  {Marinacci} F.,  {Vogelsberger}
  M.,   {Hernquist} L.,  2021b, \mn@doi [\mnras] {10.1093/mnras/stab1950},
  \href {https://ui.adsabs.harvard.edu/abs/2021MNRAS.506.4760D} {506, 4760}

\bibitem[\protect\citeauthoryear{{Dressler}}{{Dressler}}{1980}]{dressler1980}
{Dressler} A.,  1980, \mn@doi [\apj] {10.1086/157753}, \href
  {http://adsabs.harvard.edu/abs/1980ApJ...236..351D} {236, 351}

\bibitem[\protect\citeauthoryear{{Dubois} et~al.,}{{Dubois}
  et~al.}{2014}]{dubois2014}
{Dubois} Y.,  et~al., 2014, \mn@doi [\mnras] {10.1093/mnras/stu1227}, \href
  {https://ui.adsabs.harvard.edu/abs/2014MNRAS.444.1453D} {444, 1453}

\bibitem[\protect\citeauthoryear{{Dvornik} et~al.,}{{Dvornik}
  et~al.}{2020}]{dvornik2020}
{Dvornik} A.,  et~al., 2020, \mn@doi [\aap] {10.1051/0004-6361/202038693},
  \href {https://ui.adsabs.harvard.edu/abs/2020A&A...642A..83D} {642, A83}

\bibitem[\protect\citeauthoryear{{Einasto}, {Saar}, {Kaasik}  \&
  {Chernin}}{{Einasto} et~al.}{1974}]{einasto1974}
{Einasto} J.,  {Saar} E.,  {Kaasik} A.,   {Chernin} A.~D.,  1974, \mn@doi
  [\nat] {10.1038/252111a0}, \href
  {https://ui.adsabs.harvard.edu/abs/1974Natur.252..111E} {252, 111}

\bibitem[\protect\citeauthoryear{{Elahi}, {Ca{\~n}as}, {Poulton}, {Tobar},
  {Willis}, {Lagos}, {Power}  \& {Robotham}}{{Elahi} et~al.}{2019}]{elahi2019}
{Elahi} P.~J.,  {Ca{\~n}as} R.,  {Poulton} R. J.~J.,  {Tobar} R.~J.,  {Willis}
  J.~S.,  {Lagos} C. d.~P.,  {Power} C.,   {Robotham} A. S.~G.,  2019, \mn@doi
  [\pasa] {10.1017/pasa.2019.12}, \href
  {https://ui.adsabs.harvard.edu/abs/2019PASA...36...21E} {36, e021}

\bibitem[\protect\citeauthoryear{{Engler}, {Lisker}  \& {Pillepich}}{{Engler}
  et~al.}{2018}]{engler2018}
{Engler} C.,  {Lisker} T.,   {Pillepich} A.,  2018, \mn@doi [Research Notes of
  the American Astronomical Society] {10.3847/2515-5172/aabcce}, \href
  {http://adsabs.harvard.edu/abs/2018RNAAS...2b...6E} {2, 6}

\bibitem[\protect\citeauthoryear{{Erickson}, {Gottesman}  \&
  {Hunter}}{{Erickson} et~al.}{1987}]{erickson1987}
{Erickson} L.~K.,  {Gottesman} S.~T.,   {Hunter} J.~H. J.,  1987, \mn@doi
  [\nat] {10.1038/325779a0}, \href
  {https://ui.adsabs.harvard.edu/abs/1987Natur.325..779E} {325, 779}

\bibitem[\protect\citeauthoryear{{Feldmann}, {Faucher-Gigu{\`e}re}  \&
  {Kere{\v{s}}}}{{Feldmann} et~al.}{2019}]{feldmann2019}
{Feldmann} R.,  {Faucher-Gigu{\`e}re} C.-A.,   {Kere{\v{s}}} D.,  2019, \mn@doi
  [\apjl] {10.3847/2041-8213/aafe80}, \href
  {https://ui.adsabs.harvard.edu/abs/2019ApJ...871L..21F} {871, L21}

\bibitem[\protect\citeauthoryear{{Fillingham}, {Cooper}, {Pace},
  {Boylan-Kolchin}, {Bullock}, {Garrison-Kimmel}  \& {Wheeler}}{{Fillingham}
  et~al.}{2016}]{fillingham2016}
{Fillingham} S.~P.,  {Cooper} M.~C.,  {Pace} A.~B.,  {Boylan-Kolchin} M.,
  {Bullock} J.~S.,  {Garrison-Kimmel} S.,   {Wheeler} C.,  2016, \mn@doi
  [\mnras] {10.1093/mnras/stw2131}, \href
  {https://ui.adsabs.harvard.edu/\#abs/2016MNRAS.463.1916F} {463, 1916}

\bibitem[\protect\citeauthoryear{{Font} et~al.,}{{Font}
  et~al.}{2008}]{font2008}
{Font} A.~S.,  et~al., 2008, \mn@doi [\mnras]
  {10.1111/j.1365-2966.2008.13698.x}, \href
  {http://adsabs.harvard.edu/abs/2008MNRAS.389.1619F} {389, 1619}

\bibitem[\protect\citeauthoryear{{Forbes}, {Read}, {Gieles}  \&
  {Collins}}{{Forbes} et~al.}{2018}]{forbes2018}
{Forbes} D.~A.,  {Read} J.~I.,  {Gieles} M.,   {Collins} M. L.~M.,  2018,
  \mn@doi [\mnras] {10.1093/mnras/sty2584}, \href
  {https://ui.adsabs.harvard.edu/abs/2018MNRAS.481.5592F} {481, 5592}

\bibitem[\protect\citeauthoryear{{Genel} et~al.,}{{Genel}
  et~al.}{2014}]{genel2014}
{Genel} S.,  et~al., 2014, \mn@doi [\mnras] {10.1093/mnras/stu1654}, \href
  {http://adsabs.harvard.edu/abs/2014MNRAS.445..175G} {445, 175}

\bibitem[\protect\citeauthoryear{{Gnedin}, {Hernquist}  \& {Ostriker}}{{Gnedin}
  et~al.}{1999}]{gnedin1999}
{Gnedin} O.~Y.,  {Hernquist} L.,   {Ostriker} J.~P.,  1999, \mn@doi [\apj]
  {10.1086/306910}, \href
  {https://ui.adsabs.harvard.edu/abs/1999ApJ...514..109G} {514, 109}

\bibitem[\protect\citeauthoryear{{Golden-Marx} \& {Miller}}{{Golden-Marx} \&
  {Miller}}{2018}]{goldenmarx2018}
{Golden-Marx} J.~B.,  {Miller} C.~J.,  2018, \mn@doi [\apj]
  {10.3847/1538-4357/aac2bd}, \href
  {https://ui.adsabs.harvard.edu/\#abs/2018ApJ...860....2G} {860, 2}

\bibitem[\protect\citeauthoryear{{Golden-Marx} \& {Miller}}{{Golden-Marx} \&
  {Miller}}{2019}]{goldenmarx2019}
{Golden-Marx} J.~B.,  {Miller} C.~J.,  2019, \mn@doi [\apj]
  {10.3847/1538-4357/ab1d55}, \href
  {https://ui.adsabs.harvard.edu/abs/2019ApJ...878...14G} {878, 14}

\bibitem[\protect\citeauthoryear{{Grebel}}{{Grebel}}{2011}]{grebel2011}
{Grebel} E.~K.,  2011, in {Koleva} M.,  {Prugniel} P.,   {Vauglin} I.,  eds,
  EAS Publications Series Vol. 48, EAS Publications Series. pp 315--327
  (\mn@eprint {arXiv} {1103.6234}), \mn@doi{10.1051/eas/1148074}

\bibitem[\protect\citeauthoryear{{Grebel}, {Gallagher}  \& {Harbeck}}{{Grebel}
  et~al.}{2003}]{grebel2003}
{Grebel} E.~K.,  {Gallagher} J.~S.,   {Harbeck} D.,  2003, \mn@doi [\aj]
  {10.1086/368363}, \href
  {https://ui.adsabs.harvard.edu/\#abs/2003AJ....125.1926G} {125, 1926}

\bibitem[\protect\citeauthoryear{{Gu}, {Conroy}  \& {Behroozi}}{{Gu}
  et~al.}{2016}]{gu2016}
{Gu} M.,  {Conroy} C.,   {Behroozi} P.,  2016, \mn@doi [\apj]
  {10.3847/0004-637X/833/1/2}, \href
  {https://ui.adsabs.harvard.edu/\#abs/2016ApJ...833....2G} {833, 2}

\bibitem[\protect\citeauthoryear{{Gunn} \& {Gott}}{{Gunn} \&
  {Gott}}{1972}]{gunngott1972}
{Gunn} J.~E.,  {Gott} J.~Richard I.,  1972, \mn@doi [\apj] {10.1086/151605},
  \href {https://ui.adsabs.harvard.edu/\#abs/1972ApJ...176....1G} {176, 1}

\bibitem[\protect\citeauthoryear{{Guo} et~al.,}{{Guo} et~al.}{2011}]{guo2011}
{Guo} Q.,  et~al., 2011, \mn@doi [\mnras] {10.1111/j.1365-2966.2010.18114.x},
  \href {https://ui.adsabs.harvard.edu/\#abs/2011MNRAS.413..101G} {413, 101}

\bibitem[\protect\citeauthoryear{{Gupta} et~al.,}{{Gupta}
  et~al.}{2018}]{gupta2018}
{Gupta} A.,  et~al., 2018, \mn@doi [\mnras] {10.1093/mnrasl/sly037}, \href
  {http://adsabs.harvard.edu/abs/2018MNRAS.477L..35G} {477, L35}

\bibitem[\protect\citeauthoryear{{Han}, {Smith}, {Choi}, {Cortese},
  {Catinella}, {Contini}  \& {Yi}}{{Han} et~al.}{2018}]{han2018}
{Han} S.,  {Smith} R.,  {Choi} H.,  {Cortese} L.,  {Catinella} B.,  {Contini}
  E.,   {Yi} S.~K.,  2018, \mn@doi [\apj] {10.3847/1538-4357/aadfe2}, \href
  {https://ui.adsabs.harvard.edu/\#abs/2018ApJ...866...78H} {866, 78}

\bibitem[\protect\citeauthoryear{{Huang} et~al.,}{{Huang}
  et~al.}{2019}]{huang2019}
{Huang} S.,  et~al., 2019, \mn@doi [\mnras] {10.1093/mnras/stz3314}, \href
  {https://ui.adsabs.harvard.edu/abs/2019MNRAS.tmp.2990H} {p.~2990}

\bibitem[\protect\citeauthoryear{{Hudson} et~al.,}{{Hudson}
  et~al.}{2015}]{hudson2015}
{Hudson} M.~J.,  et~al., 2015, \mn@doi [\mnras] {10.1093/mnras/stu2367}, \href
  {https://ui.adsabs.harvard.edu/\#abs/2015MNRAS.447..298H} {447, 298}

\bibitem[\protect\citeauthoryear{{Jaff{\'e}} et~al.,}{{Jaff{\'e}}
  et~al.}{2018}]{jaffe2018}
{Jaff{\'e}} Y.~L.,  et~al., 2018, \mn@doi [\mnras] {10.1093/mnras/sty500},
  \href {https://ui.adsabs.harvard.edu/abs/2018MNRAS.476.4753J} {476, 4753}

\bibitem[\protect\citeauthoryear{{Joshi}, {Wadsley}  \& {Parker}}{{Joshi}
  et~al.}{2017}]{joshi2017}
{Joshi} G.~D.,  {Wadsley} J.,   {Parker} L.~C.,  2017, \mn@doi [\mnras]
  {10.1093/mnras/stx803}, \href
  {https://ui.adsabs.harvard.edu/\#abs/2017MNRAS.468.4625J} {468, 4625}

\bibitem[\protect\citeauthoryear{{Joshi}, {Parker}, {Wadsley}  \&
  {Keller}}{{Joshi} et~al.}{2019}]{joshi2019}
{Joshi} G.~D.,  {Parker} L.~C.,  {Wadsley} J.,   {Keller} B.~W.,  2019, \mn@doi
  [\mnras] {10.1093/mnras/sty3119}, \href
  {https://ui.adsabs.harvard.edu/\#abs/2019MNRAS.483..235J} {483, 235}

\bibitem[\protect\citeauthoryear{{Kawata} \& {Mulchaey}}{{Kawata} \&
  {Mulchaey}}{2008}]{kawata2008}
{Kawata} D.,  {Mulchaey} J.~S.,  2008, \mn@doi [\apjl] {10.1086/526544}, \href
  {https://ui.adsabs.harvard.edu/abs/2008ApJ...672L.103K} {672, L103}

\bibitem[\protect\citeauthoryear{{Kravtsov}, {Vikhlinin}  \&
  {Meshcheryakov}}{{Kravtsov} et~al.}{2018}]{kravtsov2018}
{Kravtsov} A.~V.,  {Vikhlinin} A.~A.,   {Meshcheryakov} A.~V.,  2018, \mn@doi
  [Astronomy Letters] {10.1134/S1063773717120015}, \href
  {https://ui.adsabs.harvard.edu/\#abs/2018AstL...44....8K} {44, 8}

\bibitem[\protect\citeauthoryear{{Larson}, {Tinsley}  \& {Caldwell}}{{Larson}
  et~al.}{1980}]{larson1980}
{Larson} R.~B.,  {Tinsley} B.~M.,   {Caldwell} C.~N.,  1980, \mn@doi [\apj]
  {10.1086/157917}, \href
  {https://ui.adsabs.harvard.edu/\#abs/1980ApJ...237..692L} {237, 692}

\bibitem[\protect\citeauthoryear{{Lewis} et~al.,}{{Lewis}
  et~al.}{2002}]{lewis2002}
{Lewis} I.,  et~al., 2002, \mn@doi [\mnras] {10.1046/j.1365-8711.2002.05558.x},
  \href {https://ui.adsabs.harvard.edu/\#abs/2002MNRAS.334..673L} {334, 673}

\bibitem[\protect\citeauthoryear{{Lin} \& {Mohr}}{{Lin} \&
  {Mohr}}{2004}]{lin2004}
{Lin} Y.-T.,  {Mohr} J.~J.,  2004, \mn@doi [\apj] {10.1086/425412}, \href
  {https://ui.adsabs.harvard.edu/abs/2004ApJ...617..879L} {617, 879}

\bibitem[\protect\citeauthoryear{{Lin}, {Mohr}  \& {Stanford}}{{Lin}
  et~al.}{2003}]{lin2003}
{Lin} Y.-T.,  {Mohr} J.~J.,   {Stanford} S.~A.,  2003, \mn@doi [\apj]
  {10.1086/375513}, \href
  {https://ui.adsabs.harvard.edu/abs/2003ApJ...591..749L} {591, 749}

\bibitem[\protect\citeauthoryear{{Lisker}, {Grebel}, {Binggeli}  \&
  {Glatt}}{{Lisker} et~al.}{2007}]{lisker2007}
{Lisker} T.,  {Grebel} E.~K.,  {Binggeli} B.,   {Glatt} K.,  2007, \mn@doi
  [\apj] {10.1086/513090}, \href
  {https://ui.adsabs.harvard.edu/\#abs/2007ApJ...660.1186L} {660, 1186}

\bibitem[\protect\citeauthoryear{{Lisker}, {Grebel}  \& {Binggeli}}{{Lisker}
  et~al.}{2008}]{lisker2008}
{Lisker} T.,  {Grebel} E.~K.,   {Binggeli} B.,  2008, \mn@doi [\aj]
  {10.1088/0004-6256/135/1/380}, \href
  {https://ui.adsabs.harvard.edu/\#abs/2008AJ....135..380L} {135, 380}

\bibitem[\protect\citeauthoryear{{Lisker}, {Weinmann}, {Janz}  \&
  {Meyer}}{{Lisker} et~al.}{2013}]{lisker2013}
{Lisker} T.,  {Weinmann} S.~M.,  {Janz} J.,   {Meyer} H.~T.,  2013, \mn@doi
  [\mnras] {10.1093/mnras/stt549}, \href
  {https://ui.adsabs.harvard.edu/\#abs/2013MNRAS.432.1162L} {432, 1162}

\bibitem[\protect\citeauthoryear{{Lisker}, {Vijayaraghavan}, {Janz},
  {Gallagher}, {Engler}  \& {Urich}}{{Lisker} et~al.}{2018}]{lisker2018}
{Lisker} T.,  {Vijayaraghavan} R.,  {Janz} J.,  {Gallagher} J.~S.,  {Engler}
  C.,   {Urich} L.,  2018, \mn@doi [\apj] {10.3847/1538-4357/aadae1}, \href
  {https://ui.adsabs.harvard.edu/\#abs/2018ApJ...865...40L} {865, 40}

\bibitem[\protect\citeauthoryear{{Mandelbaum}, {Seljak}, {Kauffmann}, {Hirata}
  \& {Brinkmann}}{{Mandelbaum} et~al.}{2006}]{mandelbaum2006}
{Mandelbaum} R.,  {Seljak} U.,  {Kauffmann} G.,  {Hirata} C.~M.,   {Brinkmann}
  J.,  2006, \mn@doi [\mnras] {10.1111/j.1365-2966.2006.10156.x}, \href
  {https://ui.adsabs.harvard.edu/abs/2006MNRAS.368..715M} {368, 715}

\bibitem[\protect\citeauthoryear{{Marinacci} et~al.,}{{Marinacci}
  et~al.}{2018}]{marinacci2018}
{Marinacci} F.,  et~al., 2018, \mn@doi [\mnras] {10.1093/mnras/sty2206}, \href
  {http://adsabs.harvard.edu/abs/2018MNRAS.480.5113M} {480, 5113}

\bibitem[\protect\citeauthoryear{{Matthee}, {Schaye}, {Crain}, {Schaller},
  {Bower}  \& {Theuns}}{{Matthee} et~al.}{2017}]{matthee2017}
{Matthee} J.,  {Schaye} J.,  {Crain} R.~A.,  {Schaller} M.,  {Bower} R.,
  {Theuns} T.,  2017, \mn@doi [\mnras] {10.1093/mnras/stw2884}, \href
  {https://ui.adsabs.harvard.edu/\#abs/2017MNRAS.465.2381M} {465, 2381}

\bibitem[\protect\citeauthoryear{{McPartland}, {Ebeling}, {Roediger}  \&
  {Blumenthal}}{{McPartland} et~al.}{2016}]{mcpartland2016}
{McPartland} C.,  {Ebeling} H.,  {Roediger} E.,   {Blumenthal} K.,  2016,
  \mn@doi [\mnras] {10.1093/mnras/stv2508}, \href
  {https://ui.adsabs.harvard.edu/abs/2016MNRAS.455.2994M} {455, 2994}

\bibitem[\protect\citeauthoryear{{Merritt}}{{Merritt}}{1983}]{merritt1983}
{Merritt} D.,  1983, \mn@doi [\apj] {10.1086/160571}, \href
  {https://ui.adsabs.harvard.edu/\#abs/1983ApJ...264...24M} {264, 24}

\bibitem[\protect\citeauthoryear{{Moore}, {Katz}, {Lake}, {Dressler}  \&
  {Oemler}}{{Moore} et~al.}{1996}]{moore1996}
{Moore} B.,  {Katz} N.,  {Lake} G.,  {Dressler} A.,   {Oemler} A.,  1996,
  \mn@doi [\nat] {10.1038/379613a0}, \href
  {https://ui.adsabs.harvard.edu/\#abs/1996Natur.379..613M} {379, 613}

\bibitem[\protect\citeauthoryear{{Moore}, {Lake}  \& {Katz}}{{Moore}
  et~al.}{1998}]{moore1998}
{Moore} B.,  {Lake} G.,   {Katz} N.,  1998, \mn@doi [\apj] {10.1086/305264},
  \href {https://ui.adsabs.harvard.edu/\#abs/1998ApJ...495..139M} {495, 139}

\bibitem[\protect\citeauthoryear{{Moster}, {Somerville}, {Maulbetsch}, {van den
  Bosch}, {Macci{\`o}}, {Naab}  \& {Oser}}{{Moster} et~al.}{2010}]{moster2010}
{Moster} B.~P.,  {Somerville} R.~S.,  {Maulbetsch} C.,  {van den Bosch} F.~C.,
  {Macci{\`o}} A.~V.,  {Naab} T.,   {Oser} L.,  2010, \mn@doi [\apj]
  {10.1088/0004-637X/710/2/903}, \href
  {http://adsabs.harvard.edu/abs/2010ApJ...710..903M} {710, 903}

\bibitem[\protect\citeauthoryear{{Moster}, {Naab}  \& {White}}{{Moster}
  et~al.}{2013}]{moster2013}
{Moster} B.~P.,  {Naab} T.,   {White} S.~D.~M.,  2013, \mn@doi [\mnras]
  {10.1093/mnras/sts261}, \href
  {http://adsabs.harvard.edu/abs/2013MNRAS.428.3121M} {428, 3121}

\bibitem[\protect\citeauthoryear{{Nagai} \& {Kravtsov}}{{Nagai} \&
  {Kravtsov}}{2005}]{nagai2005}
{Nagai} D.,  {Kravtsov} A.~V.,  2005, \mn@doi [\apj] {10.1086/426016}, \href
  {https://ui.adsabs.harvard.edu/\#abs/2005ApJ...618..557N} {618, 557}

\bibitem[\protect\citeauthoryear{{Naiman} et~al.,}{{Naiman}
  et~al.}{2018}]{naiman2018}
{Naiman} J.~P.,  et~al., 2018, \mn@doi [\mnras] {10.1093/mnras/sty618}, \href
  {http://adsabs.harvard.edu/abs/2018MNRAS.477.1206N} {477, 1206}

\bibitem[\protect\citeauthoryear{{Nelson} et~al.,}{{Nelson}
  et~al.}{2015}]{nelson2015}
{Nelson} D.,  et~al., 2015, \mn@doi [Astronomy and Computing]
  {10.1016/j.ascom.2015.09.003}, \href
  {http://adsabs.harvard.edu/abs/2015A%26C....13...12N} {13, 12}

\bibitem[\protect\citeauthoryear{{Nelson} et~al.,}{{Nelson}
  et~al.}{2018}]{nelson2018}
{Nelson} D.,  et~al., 2018, \mn@doi [\mnras] {10.1093/mnras/stx3040}, \href
  {http://adsabs.harvard.edu/abs/2018MNRAS.475..624N} {475, 624}

\bibitem[\protect\citeauthoryear{{Nelson} et~al.,}{{Nelson}
  et~al.}{2019a}]{nelson2019a}
{Nelson} D.,  et~al., 2019a, \mn@doi [Computational Astrophysics and Cosmology]
  {10.1186/s40668-019-0028-x}, \href
  {https://ui.adsabs.harvard.edu/abs/2019ComAC...6....2N} {6, 2}

\bibitem[\protect\citeauthoryear{{Nelson} et~al.,}{{Nelson}
  et~al.}{2019b}]{nelson2019b}
{Nelson} D.,  et~al., 2019b, \mn@doi [\mnras] {10.1093/mnras/stz2306}, \href
  {https://ui.adsabs.harvard.edu/abs/2019MNRAS.490.3234N} {490, 3234}

\bibitem[\protect\citeauthoryear{{Niemiec} et~al.,}{{Niemiec}
  et~al.}{2017}]{niemiec2017}
{Niemiec} A.,  et~al., 2017, \mn@doi [\mnras] {10.1093/mnras/stx1667}, \href
  {https://ui.adsabs.harvard.edu/\#abs/2017MNRAS.471.1153N} {471, 1153}

\bibitem[\protect\citeauthoryear{{Niemiec}, {Jullo}, {Giocoli}, {Limousin}  \&
  {Jauzac}}{{Niemiec} et~al.}{2019}]{niemiec2019}
{Niemiec} A.,  {Jullo} E.,  {Giocoli} C.,  {Limousin} M.,   {Jauzac} M.,  2019,
  \mn@doi [\mnras] {10.1093/mnras/stz1318}, \href
  {https://ui.adsabs.harvard.edu/abs/2019MNRAS.487..653N} {487, 653}

\bibitem[\protect\citeauthoryear{{Oemler}}{{Oemler}}{1974}]{oemler1974}
{Oemler} Augustus J.,  1974, \mn@doi [\apj] {10.1086/153216}, \href
  {https://ui.adsabs.harvard.edu/abs/1974ApJ...194....1O} {194, 1}

\bibitem[\protect\citeauthoryear{{Pakmor} \& {Springel}}{{Pakmor} \&
  {Springel}}{2013}]{pakmor2013}
{Pakmor} R.,  {Springel} V.,  2013, \mn@doi [\mnras] {10.1093/mnras/stt428},
  \href {http://adsabs.harvard.edu/abs/2013MNRAS.432..176P} {432, 176}

\bibitem[\protect\citeauthoryear{{Pasquali}, {Gallazzi}, {Fontanot}, {van den
  Bosch}, {De Lucia}, {Mo}  \& {Yang}}{{Pasquali} et~al.}{2010}]{pasquali2010}
{Pasquali} A.,  {Gallazzi} A.,  {Fontanot} F.,  {van den Bosch} F.~C.,  {De
  Lucia} G.,  {Mo} H.~J.,   {Yang} X.,  2010, \mn@doi [\mnras]
  {10.1111/j.1365-2966.2010.17074.x}, \href
  {http://adsabs.harvard.edu/abs/2010MNRAS.407..937P} {407, 937}

\bibitem[\protect\citeauthoryear{{Pasquali}, {Smith}, {Gallazzi}, {De Lucia},
  {Zibetti}, {Hirschmann}  \& {Yi}}{{Pasquali} et~al.}{2019}]{pasquali2019}
{Pasquali} A.,  {Smith} R.,  {Gallazzi} A.,  {De Lucia} G.,  {Zibetti} S.,
  {Hirschmann} M.,   {Yi} S.~K.,  2019, \mn@doi [\mnras]
  {10.1093/mnras/sty3530}, \href
  {https://ui.adsabs.harvard.edu/\#abs/2019MNRAS.484.1702P} {484, 1702}

\bibitem[\protect\citeauthoryear{{Peng}, {Ford}  \& {Freeman}}{{Peng}
  et~al.}{2004}]{peng2004}
{Peng} E.~W.,  {Ford} H.~C.,   {Freeman} K.~C.,  2004, \mn@doi [\apj]
  {10.1086/381160}, \href
  {https://ui.adsabs.harvard.edu/abs/2004ApJ...602..685P} {602, 685}

\bibitem[\protect\citeauthoryear{{Peng} et~al.,}{{Peng}
  et~al.}{2010}]{peng2010}
{Peng} Y.-j.,  et~al., 2010, \mn@doi [\apj] {10.1088/0004-637X/721/1/193},
  \href {https://ui.adsabs.harvard.edu/abs/2010ApJ...721..193P} {721, 193}

\bibitem[\protect\citeauthoryear{{Pillepich} et~al.,}{{Pillepich}
  et~al.}{2018a}]{pillepich2018a}
{Pillepich} A.,  et~al., 2018a, \mn@doi [\mnras] {10.1093/mnras/stx2656}, \href
  {http://adsabs.harvard.edu/abs/2018MNRAS.473.4077P} {473, 4077}

\bibitem[\protect\citeauthoryear{{Pillepich} et~al.,}{{Pillepich}
  et~al.}{2018b}]{pillepich2018b}
{Pillepich} A.,  et~al., 2018b, \mn@doi [\mnras] {10.1093/mnras/stx3112}, \href
  {http://adsabs.harvard.edu/abs/2018MNRAS.475..648P} {475, 648}

\bibitem[\protect\citeauthoryear{{Pillepich} et~al.,}{{Pillepich}
  et~al.}{2019}]{pillepich2019}
{Pillepich} A.,  et~al., 2019, \mn@doi [\mnras] {10.1093/mnras/stz2338}, \href
  {https://ui.adsabs.harvard.edu/abs/2019MNRAS.490.3196P} {490, 3196}

\bibitem[\protect\citeauthoryear{{Planck Collaboration} et~al.,}{{Planck
  Collaboration} et~al.}{2016}]{planck2016}
{Planck Collaboration} et~al., 2016, \mn@doi [\aap]
  {10.1051/0004-6361/201527101}, \href
  {https://ui.adsabs.harvard.edu/#abs/2016A&A...594A...1P} {594, A1}

\bibitem[\protect\citeauthoryear{{Prescott} et~al.,}{{Prescott}
  et~al.}{2011}]{prescott2011}
{Prescott} M.,  et~al., 2011, \mn@doi [\mnras]
  {10.1111/j.1365-2966.2011.19353.x}, \href
  {https://ui.adsabs.harvard.edu/\#abs/2011MNRAS.417.1374P} {417, 1374}

\bibitem[\protect\citeauthoryear{{Prole} et~al.,}{{Prole}
  et~al.}{2019}]{prole2019}
{Prole} D.~J.,  et~al., 2019, \mn@doi [\mnras] {10.1093/mnras/stz326}, \href
  {https://ui.adsabs.harvard.edu/abs/2019MNRAS.484.4865P} {484, 4865}

\bibitem[\protect\citeauthoryear{{Reddick}, {Wechsler}, {Tinker}  \&
  {Behroozi}}{{Reddick} et~al.}{2013}]{reddick2013}
{Reddick} R.~M.,  {Wechsler} R.~H.,  {Tinker} J.~L.,   {Behroozi} P.~S.,  2013,
  \mn@doi [\apj] {10.1088/0004-637X/771/1/30}, \href
  {https://ui.adsabs.harvard.edu/\#abs/2013ApJ...771...30R} {771, 30}

\bibitem[\protect\citeauthoryear{{Rhee}, {Smith}, {Choi}, {Yi}, {Jaff{\'e}},
  {Candlish}  \& {S{\'a}nchez-J{\'a}nssen}}{{Rhee} et~al.}{2017}]{rhee2017}
{Rhee} J.,  {Smith} R.,  {Choi} H.,  {Yi} S.~K.,  {Jaff{\'e}} Y.,  {Candlish}
  G.,   {S{\'a}nchez-J{\'a}nssen} R.,  2017, \mn@doi [\apj]
  {10.3847/1538-4357/aa6d6c}, \href
  {https://ui.adsabs.harvard.edu/\#abs/2017ApJ...843..128R} {843, 128}

\bibitem[\protect\citeauthoryear{{Rodr{\'\i}guez-Puebla}, {Drory}  \&
  {Avila-Reese}}{{Rodr{\'\i}guez-Puebla} et~al.}{2012}]{rodriguezPuebla2012}
{Rodr{\'\i}guez-Puebla} A.,  {Drory} N.,   {Avila-Reese} V.,  2012, \mn@doi
  [\apj] {10.1088/0004-637X/756/1/2}, \href
  {https://ui.adsabs.harvard.edu/abs/2012ApJ...756....2R} {756, 2}

\bibitem[\protect\citeauthoryear{{Rodr{\'\i}guez-Puebla}, {Avila-Reese}  \&
  {Drory}}{{Rodr{\'\i}guez-Puebla} et~al.}{2013}]{rodriguezPuebla2013}
{Rodr{\'\i}guez-Puebla} A.,  {Avila-Reese} V.,   {Drory} N.,  2013, \mn@doi
  [\apj] {10.1088/0004-637X/767/1/92}, \href
  {https://ui.adsabs.harvard.edu/abs/2013ApJ...767...92R} {767, 92}

\bibitem[\protect\citeauthoryear{{Safarzadeh} \& {Loeb}}{{Safarzadeh} \&
  {Loeb}}{2019}]{safarzadeh2019}
{Safarzadeh} M.,  {Loeb} A.,  2019, \mn@doi [\mnras] {10.1093/mnrasl/slz053},
  \href {https://ui.adsabs.harvard.edu/abs/2019MNRAS.486L..26S} {486, L26}

\bibitem[\protect\citeauthoryear{{Sales} et~al.,}{{Sales}
  et~al.}{2015}]{sales2015}
{Sales} L.~V.,  et~al., 2015, \mn@doi [\mnras] {10.1093/mnrasl/slu173}, \href
  {https://ui.adsabs.harvard.edu/\#abs/2015MNRAS.447L...6S} {447, L6}

\bibitem[\protect\citeauthoryear{{Schaye} et~al.,}{{Schaye}
  et~al.}{2015}]{schaye2015}
{Schaye} J.,  et~al., 2015, \mn@doi [\mnras] {10.1093/mnras/stu2058}, \href
  {https://ui.adsabs.harvard.edu/\#abs/2015MNRAS.446..521S} {446, 521}

\bibitem[\protect\citeauthoryear{{Shi} et~al.,}{{Shi} et~al.}{2020}]{shi2020}
{Shi} J.,  et~al., 2020, \mn@doi [\apj] {10.3847/1538-4357/ab8464}, \href
  {https://ui.adsabs.harvard.edu/abs/2020ApJ...893..139S} {893, 139}

\bibitem[\protect\citeauthoryear{{Sif{\'o}n}, {Herbonnet}, {Hoekstra}, {van der
  Burg}  \& {Viola}}{{Sif{\'o}n} et~al.}{2018}]{sifon2018}
{Sif{\'o}n} C.,  {Herbonnet} R.,  {Hoekstra} H.,  {van der Burg} R. F.~J.,
  {Viola} M.,  2018, \mn@doi [\mnras] {10.1093/mnras/sty1161}, \href
  {https://ui.adsabs.harvard.edu/\#abs/2018MNRAS.478.1244S} {478, 1244}

\bibitem[\protect\citeauthoryear{{Sijacki}, {Vogelsberger}, {Genel},
  {Springel}, {Torrey}, {Snyder}, {Nelson}  \& {Hernquist}}{{Sijacki}
  et~al.}{2015}]{sijacki2015}
{Sijacki} D.,  {Vogelsberger} M.,  {Genel} S.,  {Springel} V.,  {Torrey} P.,
  {Snyder} G.~F.,  {Nelson} D.,   {Hernquist} L.,  2015, \mn@doi [\mnras]
  {10.1093/mnras/stv1340}, \href
  {http://adsabs.harvard.edu/abs/2015MNRAS.452..575S} {452, 575}

\bibitem[\protect\citeauthoryear{{Simpson}, {Grand}, {G{\'o}mez}, {Marinacci},
  {Pakmor}, {Springel}, {Campbell}  \& {Frenk}}{{Simpson}
  et~al.}{2018}]{simpson2018}
{Simpson} C.~M.,  {Grand} R. J.~J.,  {G{\'o}mez} F.~A.,  {Marinacci} F.,
  {Pakmor} R.,  {Springel} V.,  {Campbell} D. J.~R.,   {Frenk} C.~S.,  2018,
  \mn@doi [\mnras] {10.1093/mnras/sty774}, \href
  {https://ui.adsabs.harvard.edu/abs/2018MNRAS.478..548S} {478, 548}

\bibitem[\protect\citeauthoryear{{Smith}, {S{\'a}nchez-Janssen}, {Fellhauer},
  {Puzia}, {Aguerri}  \& {Farias}}{{Smith} et~al.}{2013}]{smith2013}
{Smith} R.,  {S{\'a}nchez-Janssen} R.,  {Fellhauer} M.,  {Puzia} T.~H.,
  {Aguerri} J.~A.~L.,   {Farias} J.~P.,  2013, \mn@doi [\mnras]
  {10.1093/mnras/sts395}, \href
  {https://ui.adsabs.harvard.edu/\#abs/2013MNRAS.429.1066S} {429, 1066}

\bibitem[\protect\citeauthoryear{{Smith} et~al.,}{{Smith}
  et~al.}{2015}]{smith2015}
{Smith} R.,  et~al., 2015, \mn@doi [\mnras] {10.1093/mnras/stv2082}, \href
  {https://ui.adsabs.harvard.edu/\#abs/2015MNRAS.454.2502S} {454, 2502}

\bibitem[\protect\citeauthoryear{{Smith}, {Choi}, {Lee}, {Rhee},
  {Sanchez-Janssen}  \& {Yi}}{{Smith} et~al.}{2016}]{smith2016}
{Smith} R.,  {Choi} H.,  {Lee} J.,  {Rhee} J.,  {Sanchez-Janssen} R.,   {Yi}
  S.~K.,  2016, \mn@doi [\apj] {10.3847/1538-4357/833/1/109}, \href
  {https://ui.adsabs.harvard.edu/\#abs/2016ApJ...833..109S} {833, 109}

\bibitem[\protect\citeauthoryear{{Snyder} et~al.,}{{Snyder}
  et~al.}{2015}]{snyder2015}
{Snyder} G.~F.,  et~al., 2015, \mn@doi [\mnras] {10.1093/mnras/stv2078}, \href
  {https://ui.adsabs.harvard.edu/abs/2015MNRAS.454.1886S} {454, 1886}

\bibitem[\protect\citeauthoryear{{Sonnenfeld}, {Wang}  \&
  {Bahcall}}{{Sonnenfeld} et~al.}{2019}]{sonnenfeld2019}
{Sonnenfeld} A.,  {Wang} W.,   {Bahcall} N.,  2019, \mn@doi [\aap]
  {10.1051/0004-6361/201834260}, \href
  {https://ui.adsabs.harvard.edu/\#abs/2019A&A...622A..30S} {622, A30}

\bibitem[\protect\citeauthoryear{{Spindler} et~al.,}{{Spindler}
  et~al.}{2018}]{spindler2018}
{Spindler} A.,  et~al., 2018, \mn@doi [\mnras] {10.1093/mnras/sty247}, \href
  {https://ui.adsabs.harvard.edu/\#abs/2018MNRAS.476..580S} {476, 580}

\bibitem[\protect\citeauthoryear{{Spitler} \& {Forbes}}{{Spitler} \&
  {Forbes}}{2009}]{spitler2009}
{Spitler} L.~R.,  {Forbes} D.~A.,  2009, \mn@doi [\mnras]
  {10.1111/j.1745-3933.2008.00567.x}, \href
  {https://ui.adsabs.harvard.edu/abs/2009MNRAS.392L...1S} {392, L1}

\bibitem[\protect\citeauthoryear{{Springel}}{{Springel}}{2010}]{springel2010}
{Springel} V.,  2010, \mn@doi [\mnras] {10.1111/j.1365-2966.2009.15715.x},
  \href {http://adsabs.harvard.edu/abs/2010MNRAS.401..791S} {401, 791}

\bibitem[\protect\citeauthoryear{{Springel} \& {Hernquist}}{{Springel} \&
  {Hernquist}}{2003}]{springel2003}
{Springel} V.,  {Hernquist} L.,  2003, \mn@doi [\mnras]
  {10.1046/j.1365-8711.2003.06206.x}, \href
  {https://ui.adsabs.harvard.edu/abs/2003MNRAS.339..289S} {339, 289}

\bibitem[\protect\citeauthoryear{{Springel}, {White}, {Tormen}  \&
  {Kauffmann}}{{Springel} et~al.}{2001}]{springel2001}
{Springel} V.,  {White} S.~D.~M.,  {Tormen} G.,   {Kauffmann} G.,  2001,
  \mn@doi [\mnras] {10.1046/j.1365-8711.2001.04912.x}, \href
  {http://adsabs.harvard.edu/abs/2001MNRAS.328..726S} {328, 726}

\bibitem[\protect\citeauthoryear{{Springel} et~al.,}{{Springel}
  et~al.}{2005}]{springel2005}
{Springel} V.,  et~al., 2005, \mn@doi [\nat] {10.1038/nature03597}, \href
  {https://ui.adsabs.harvard.edu/\#abs/2005Natur.435..629S} {435, 629}

\bibitem[\protect\citeauthoryear{{Springel} et~al.,}{{Springel}
  et~al.}{2018}]{springel2018}
{Springel} V.,  et~al., 2018, \mn@doi [\mnras] {10.1093/mnras/stx3304}, \href
  {http://adsabs.harvard.edu/abs/2018MNRAS.475..676S} {475, 676}

\bibitem[\protect\citeauthoryear{{Sybilska} et~al.,}{{Sybilska}
  et~al.}{2017}]{sybilska2017}
{Sybilska} A.,  et~al., 2017, \mn@doi [\mnras] {10.1093/mnras/stx1138}, \href
  {http://adsabs.harvard.edu/abs/2017MNRAS.470..815S} {470, 815}

\bibitem[\protect\citeauthoryear{{Terrazas} et~al.,}{{Terrazas}
  et~al.}{2020}]{terrazas2020}
{Terrazas} B.~A.,  et~al., 2020, \mn@doi [\mnras] {10.1093/mnras/staa374},
  \href {https://ui.adsabs.harvard.edu/abs/2020MNRAS.tmp..423T} {}

\bibitem[\protect\citeauthoryear{{Tinker}, {Leauthaud}, {Bundy}, {George},
  {Behroozi}, {Massey}, {Rhodes}  \& {Wechsler}}{{Tinker}
  et~al.}{2013}]{tinker2013}
{Tinker} J.~L.,  {Leauthaud} A.,  {Bundy} K.,  {George} M.~R.,  {Behroozi} P.,
  {Massey} R.,  {Rhodes} J.,   {Wechsler} R.~H.,  2013, \mn@doi [\apj]
  {10.1088/0004-637X/778/2/93}, \href
  {https://ui.adsabs.harvard.edu/\#abs/2013ApJ...778...93T} {778, 93}

\bibitem[\protect\citeauthoryear{{Tonnesen} \& {Cen}}{{Tonnesen} \&
  {Cen}}{2015}]{tonnesen2015}
{Tonnesen} S.,  {Cen} R.,  2015, \mn@doi [\apj] {10.1088/0004-637X/812/2/104},
  \href {https://ui.adsabs.harvard.edu/\#abs/2015ApJ...812..104T} {812, 104}

\bibitem[\protect\citeauthoryear{{Tonnesen}, {Bryan}  \& {van
  Gorkom}}{{Tonnesen} et~al.}{2007}]{tonnesen2007}
{Tonnesen} S.,  {Bryan} G.~L.,   {van Gorkom} J.~H.,  2007, \mn@doi [\apj]
  {10.1086/523034}, \href
  {https://ui.adsabs.harvard.edu/abs/2007ApJ...671.1434T} {671, 1434}

\bibitem[\protect\citeauthoryear{{Torrey}, {Vogelsberger}, {Genel}, {Sijacki},
  {Springel}  \& {Hernquist}}{{Torrey} et~al.}{2014}]{torrey2014}
{Torrey} P.,  {Vogelsberger} M.,  {Genel} S.,  {Sijacki} D.,  {Springel} V.,
  {Hernquist} L.,  2014, \mn@doi [\mnras] {10.1093/mnras/stt2295}, \href
  {http://adsabs.harvard.edu/abs/2014MNRAS.438.1985T} {438, 1985}

\bibitem[\protect\citeauthoryear{{Tully} \& {Fisher}}{{Tully} \&
  {Fisher}}{1977}]{tully1977}
{Tully} R.~B.,  {Fisher} J.~R.,  1977, \aap, \href
  {https://ui.adsabs.harvard.edu/abs/1977A&A....54..661T} {500, 105}

\bibitem[\protect\citeauthoryear{{Vijayaraghavan} \& {Ricker}}{{Vijayaraghavan}
  \& {Ricker}}{2013}]{vijayaraghavan2013}
{Vijayaraghavan} R.,  {Ricker} P.~M.,  2013, \mn@doi [\mnras]
  {10.1093/mnras/stt1485}, \href
  {https://ui.adsabs.harvard.edu/\#abs/2013MNRAS.435.2713V} {435, 2713}

\bibitem[\protect\citeauthoryear{{Vogelsberger}, {Genel}, {Sijacki}, {Torrey},
  {Springel}  \& {Hernquist}}{{Vogelsberger} et~al.}{2013}]{vogelsberger2013}
{Vogelsberger} M.,  {Genel} S.,  {Sijacki} D.,  {Torrey} P.,  {Springel} V.,
  {Hernquist} L.,  2013, \mn@doi [\mnras] {10.1093/mnras/stt1789}, \href
  {http://adsabs.harvard.edu/abs/2013MNRAS.436.3031V} {436, 3031}

\bibitem[\protect\citeauthoryear{{Vogelsberger} et~al.,}{{Vogelsberger}
  et~al.}{2014a}]{vogelsberger2014b}
{Vogelsberger} M.,  et~al., 2014a, \mn@doi [\mnras] {10.1093/mnras/stu1536},
  \href {http://adsabs.harvard.edu/abs/2014MNRAS.444.1518V} {444, 1518}

\bibitem[\protect\citeauthoryear{{Vogelsberger} et~al.,}{{Vogelsberger}
  et~al.}{2014b}]{vogelsberger2014a}
{Vogelsberger} M.,  et~al., 2014b, \mn@doi [\nat] {10.1038/nature13316}, \href
  {http://adsabs.harvard.edu/abs/2014Natur.509..177V} {509, 177}

\bibitem[\protect\citeauthoryear{{Vogelsberger} et~al.,}{{Vogelsberger}
  et~al.}{2018}]{vogelsberger2018}
{Vogelsberger} M.,  et~al., 2018, \mn@doi [\mnras] {10.1093/mnras/stx2955},
  \href {https://ui.adsabs.harvard.edu/abs/2018MNRAS.474.2073V} {474, 2073}

\bibitem[\protect\citeauthoryear{{Vogelsberger} et~al.,}{{Vogelsberger}
  et~al.}{2020}]{vogelsberger2020}
{Vogelsberger} M.,  et~al., 2020, \mn@doi [\mnras] {10.1093/mnras/staa137},
  \href {https://ui.adsabs.harvard.edu/abs/2020MNRAS.492.5167V} {492, 5167}

\bibitem[\protect\citeauthoryear{{Vulcani} et~al.,}{{Vulcani}
  et~al.}{2018}]{vulcani2018}
{Vulcani} B.,  et~al., 2018, \mn@doi [\apjl] {10.3847/2041-8213/aae68b}, \href
  {https://ui.adsabs.harvard.edu/abs/2018ApJ...866L..25V} {866, L25}

\bibitem[\protect\citeauthoryear{{Weinberger} et~al.,}{{Weinberger}
  et~al.}{2017}]{weinberger2017}
{Weinberger} R.,  et~al., 2017, \mn@doi [\mnras] {10.1093/mnras/stw2944}, \href
  {http://adsabs.harvard.edu/abs/2017MNRAS.465.3291W} {465, 3291}

\bibitem[\protect\citeauthoryear{{Weinmann}, {Lisker}, {Guo}, {Meyer}  \&
  {Janz}}{{Weinmann} et~al.}{2011}]{weinmann2011}
{Weinmann} S.~M.,  {Lisker} T.,  {Guo} Q.,  {Meyer} H.~T.,   {Janz} J.,  2011,
  \mn@doi [\mnras] {10.1111/j.1365-2966.2011.19118.x}, \href
  {http://adsabs.harvard.edu/abs/2011MNRAS.416.1197W} {416, 1197}

\bibitem[\protect\citeauthoryear{{Wetzel}, {Tinker}, {Conroy}  \& {van den
  Bosch}}{{Wetzel} et~al.}{2013}]{wetzel2013}
{Wetzel} A.~R.,  {Tinker} J.~L.,  {Conroy} C.,   {van den Bosch} F.~C.,  2013,
  \mn@doi [\mnras] {10.1093/mnras/stt469}, \href
  {https://ui.adsabs.harvard.edu/abs/2013MNRAS.432..336W} {432, 336}

\bibitem[\protect\citeauthoryear{{Yang}, {Mo}, {van den Bosch}, {Pasquali},
  {Li}  \& {Barden}}{{Yang} et~al.}{2007}]{yang2007}
{Yang} X.,  {Mo} H.~J.,  {van den Bosch} F.~C.,  {Pasquali} A.,  {Li} C.,
  {Barden} M.,  2007, \mn@doi [\apj] {10.1086/522027}, \href
  {https://ui.adsabs.harvard.edu/abs/2007ApJ...671..153Y} {671, 153}

\bibitem[\protect\citeauthoryear{{Yun} et~al.,}{{Yun} et~al.}{2019}]{yun2019}
{Yun} K.,  et~al., 2019, \mn@doi [\mnras] {10.1093/mnras/sty3156}, \href
  {https://ui.adsabs.harvard.edu/abs/2019MNRAS.483.1042Y} {483, 1042}

\bibitem[\protect\citeauthoryear{{Zinger}, {Dekel}, {Kravtsov}  \&
  {Nagai}}{{Zinger} et~al.}{2018}]{zinger2018}
{Zinger} E.,  {Dekel} A.,  {Kravtsov} A.~V.,   {Nagai} D.,  2018, \mn@doi
  [\mnras] {10.1093/mnras/stx3329}, \href
  {https://ui.adsabs.harvard.edu/\#abs/2018MNRAS.475.3654Z} {475, 3654}

\bibitem[\protect\citeauthoryear{{Zinger} et~al.,}{{Zinger}
  et~al.}{2020}]{zinger2020}
{Zinger} E.,  et~al., 2020, \mn@doi [\mnras] {10.1093/mnras/staa2607}, \href
  {https://ui.adsabs.harvard.edu/abs/2020MNRAS.tmp.2034Z} {}

\bibitem[\protect\citeauthoryear{{van Uitert} et~al.,}{{van Uitert}
  et~al.}{2016}]{vanuitert2016}
{van Uitert} E.,  et~al., 2016, \mn@doi [\mnras] {10.1093/mnras/stw747}, \href
  {https://ui.adsabs.harvard.edu/\#abs/2016MNRAS.459.3251V} {459, 3251}

\bibitem[\protect\citeauthoryear{{van den Bosch} \& {Ogiya}}{{van den Bosch} \&
  {Ogiya}}{2018}]{vandenbosch2018}
{van den Bosch} F.~C.,  {Ogiya} G.,  2018, \mn@doi [\mnras]
  {10.1093/mnras/sty084}, \href
  {https://ui.adsabs.harvard.edu/abs/2018MNRAS.475.4066V} {475, 4066}

\bibitem[\protect\citeauthoryear{{van den Bosch}, {Norberg}, {Mo}  \&
  {Yang}}{{van den Bosch} et~al.}{2004}]{vandenbosch2004}
{van den Bosch} F.~C.,  {Norberg} P.,  {Mo} H.~J.,   {Yang} X.,  2004, \mn@doi
  [\mnras] {10.1111/j.1365-2966.2004.08021.x}, \href
  {https://ui.adsabs.harvard.edu/abs/2004MNRAS.352.1302V} {352, 1302}

\bibitem[\protect\citeauthoryear{{van den Bosch}, {Aquino}, {Yang}, {Mo},
  {Pasquali}, {McIntosh}, {Weinmann}  \& {Kang}}{{van den Bosch}
  et~al.}{2008}]{vandenbosch2008}
{van den Bosch} F.~C.,  {Aquino} D.,  {Yang} X.,  {Mo} H.~J.,  {Pasquali} A.,
  {McIntosh} D.~H.,  {Weinmann} S.~M.,   {Kang} X.,  2008, \mn@doi [\mnras]
  {10.1111/j.1365-2966.2008.13230.x}, \href
  {https://ui.adsabs.harvard.edu/abs/2008MNRAS.387...79V} {387, 79}

\bibitem[\protect\citeauthoryear{{van der Wel}, {Bell}, {Holden}, {Skibba}  \&
  {Rix}}{{van der Wel} et~al.}{2010}]{vanderwel2010}
{van der Wel} A.,  {Bell} E.~F.,  {Holden} B.~P.,  {Skibba} R.~A.,   {Rix}
  H.-W.,  2010, \mn@doi [\apj] {10.1088/0004-637X/714/2/1779}, \href
  {https://ui.adsabs.harvard.edu/\#abs/2010ApJ...714.1779V} {714, 1779}

\bibitem[\protect\citeauthoryear{{von der Linden}, {Wild}, {Kauffmann}, {White}
   \& {Weinmann}}{{von der Linden} et~al.}{2010}]{vonderlinden2010}
{von der Linden} A.,  {Wild} V.,  {Kauffmann} G.,  {White} S. D.~M.,
  {Weinmann} S.,  2010, \mn@doi [\mnras] {10.1111/j.1365-2966.2010.16375.x},
  \href {https://ui.adsabs.harvard.edu/abs/2010MNRAS.404.1231V} {404, 1231}

\makeatother
\end{thebibliography}



\appendix

\section{Rescaling stellar mass}
\label{sec:resc}

\begin{figure*}
    \centering
    \includegraphics[width=.88\textwidth]{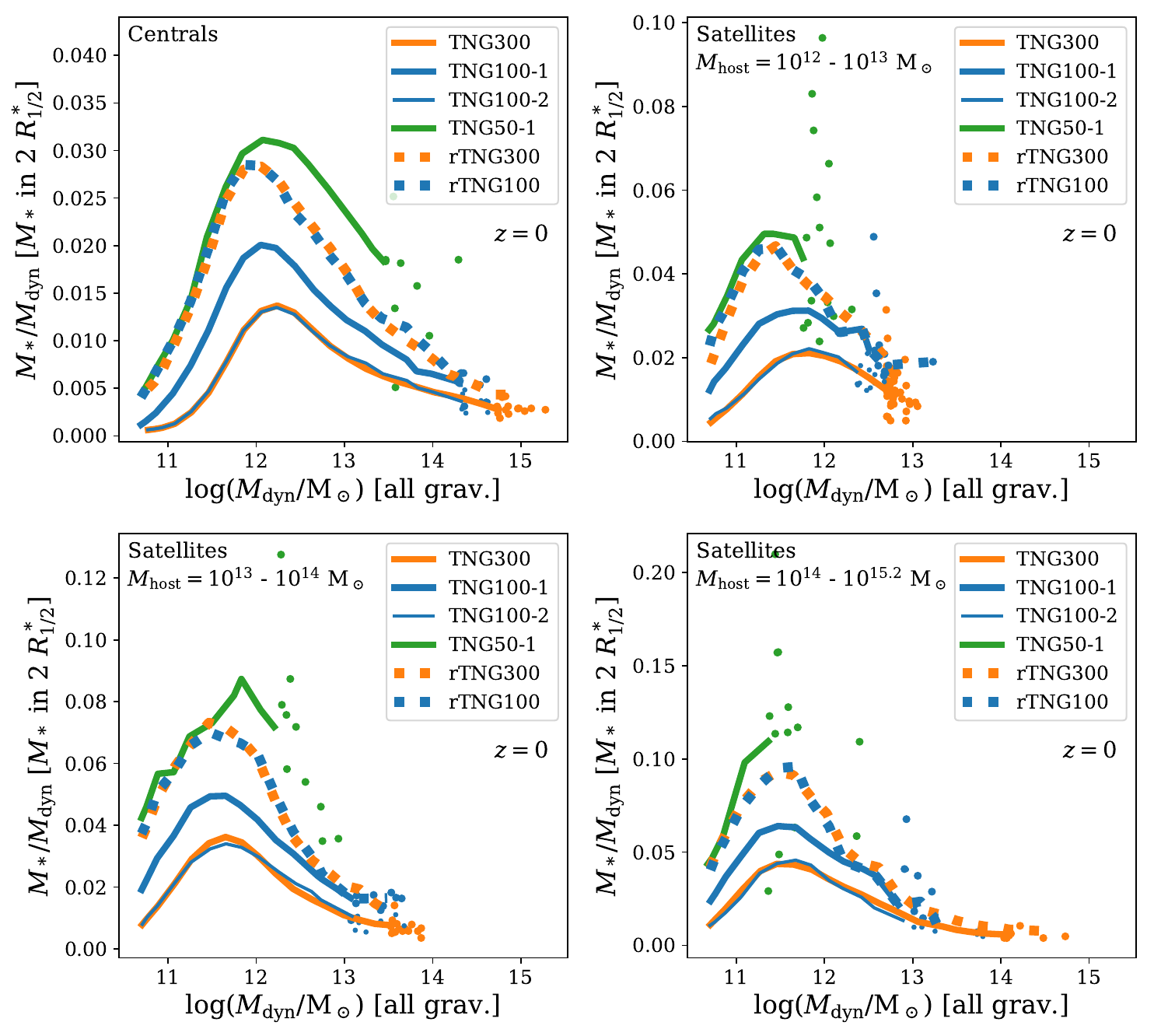}
    \caption{Stellar-to-halo mass relations at $z=0$ within fiducial apertures, illustrating the deviations between simulation volumes and resolution levels (TNG300 in orange, TNG100 and TNG100-2 in blue, TNG50 in green). We rescale stellar masses in TNG300 and TNG100 with respect to their host halo -- i.e. separately for centrals (top left) and satellites in hosts of $10^{12}-10^{13}~\rmn{M}_\odot$ (top right), $10^{13}-10^{14}~\rmn{M}_\odot$ hosts (bottom left), and $10^{14}-10^{15.2}~\rmn{M}_\odot$ (bottom right). Dotted orange and blue curves depict rescaled stellar masses for rTNG300 and rTNG100, i.e. resolution-correct values (see text for details).} 
    \label{fig:rescale}
\end{figure*}

In order to combine the statistics available in TNG300 with the improved resolution of TNG100 and TNG50, we rescale stellar masses as a function of dynamical mass by utilising the differences between simulation volumes and resolution levels. For similar approaches and motivations we refer the reader to \cite{pillepich2018b} and \cite{vogelsberger2018, vogelsberger2020}. 

While TNG100 (aka TNG100-1) was run at a baryonic mass resolution of $1.4 \times10^{6}~\rmn{M}_\odot$, both TNG100-2 (the lower resolution version of TNG100) and TNG300 employ a mass resolution lower by a factor of 8 at $1.1 \times 10^{7}~\rmn{M}_\odot$. TNG50 reaches a mass resolution of $8.5 \times 10^{4}~\rmn{M}_\odot$ -- higher than TNG100-1 by a factor of 16 (see also Table~\ref{tab:sims} for more details on differences between simulation runs). 

Figure~\ref{fig:rescale} illustrates the resolution effects on the SHMR as stellar mass fractions as a function of dynamical mass for centrals (upper left panel), as well as satellites in hosts of $10^{12} - 10^{13}~\rmn{M}_\odot$ (top right), $10^{13} - 10^{14}~\rmn{M}_\odot$ (bottom left), and $10^{14} - 10^{15.2}~\rmn{M}_\odot$ (bottom right). Solid curves correspond to the original SHMRs of different simulation runs: TNG300 (orange curve), TNG100-1 and TNG100-2 (thick and thin blue curves), as well as TNG50 (green curve). It is reassuring that, despite the different volume realizations and sizes, the outputs of TNG100-2 and TNG300 are perfectly consistent.

Now, dotted curves depict the {\it rescaled} SHMRs for rTNG300 (orange) and rTNG100 (blue), namely the resolution-corrected values with the same effective numerical mass resolution as in TNG50. In the following, we give more details on the procedure we adopt to obtain them.

Overall, we follow the approach in \cite{pillepich2018b}. However, differently from there, since centrals and satellites in different hosts form distinct SHMRs, we rescale them separately according to their environment. However, due to the statistics available, we only rescale stellar masses to TNG50 at low to intermediate dynamical masses. At higher dynamical mass, we switch to TNG100 as a reference, by in practice following a two-step procedure. Firstly, for rTNG300, we utilize the offset between TNG100-1 and TNG100-2 to rescale the stellar masses in TNG300 to the resolution of TNG100:

\begin{equation}
\begin{split}
    M_* (&M_\rmn{dyn}; \text{(r)TNG300}) \\&= M_*(M_\rmn{dyn}, \text{TNG300}) \cdot \frac{M_*(M_\rmn{dyn}, \text{TNG100-1})}{M_*(M_\rmn{dyn}, \text{TNG100-2})}.
    \label{eq:rescale_toTNG100}
\end{split}
\end{equation}

Here, $M_*(M_\rmn{dyn}, \text{TNG100-1})$ and $M_*(M_\rmn{dyn}, \text{TNG100-2})$ correspond to the average stellar mass at the respective dynamical mass and resolution level. We apply this scaling in bins of total dynamical mass to an upper limit. For centrals, this corresponds to $M_\rmn{dyn} = 10^{14}~\rmn{M}_\odot$; for more massive centrals, the fraction in Equation~\eqref{eq:rescale_toTNG100} is averaged for all centrals with $M_\rmn{dyn} = 10^{13} - 10^{14}~\rmn{M}_\odot$. We proceed similarly with satellites: satellites in $10^{12} - 10^{13}~\rmn{M}_\odot$ hosts are rescaled according to Equation~\eqref{eq:rescale_toTNG100} up to $M_\rmn{dyn} = 10^{12}~\rmn{M}_\odot$, for more massive satellites the rescaling factor is averaged over all satellites with $M_\rmn{dyn} \geq 10^{12.5}~\rmn{M}_\odot$. For satellites in $10^{13} - 10^{14}~\rmn{M}_\odot$ and $10^{14} - 10^{15.2}~\rmn{M}_\odot$ hosts, we apply an upper limit of $10^{13.5}~\rmn{M}_\odot$ and $10^{13}~\rmn{M}_\odot$. In order to rescale the massive end, we use the average rescaling factor for satellites with $M_\rmn{dyn} \geq 10^{12.5}~\rmn{M}_\odot$ in both host mass bins.

Finally, we rescale stellar masses for galaxies at lower and intermediate dynamical masses of both TNG300 and TNG100-1 to TNG50, according to the offset between TNG50 and TNG100-1:

\begin{align}
\begin{split}
M_* (&M_\rmn{dyn}; \text{rTNG300}) \\&= M_*(M_\rmn{dyn}; \text{(r)TNG300}) \cdot \frac{M_*(M_\rmn{dyn}; \text{TNG50})}{M_*(M_\rmn{dyn}; \text{TNG100-1})}, \label{eq:rescaleTNG300_toTNG50}
\end{split}\\
\begin{split}
M_* (&M_\rmn{dyn}; \text{rTNG100}) \\&= M_*(M_\rmn{dyn}; \text{TNG100}) \cdot \frac{M_*(M_\rmn{dyn}; \text{TNG50})}{M_*(M_\rmn{dyn}; \text{TNG100-1})}.
\end{split}
\label{eq:rescaleTGN100_toTNG50}
\end{align}

As in Equation~\eqref{eq:rescale_toTNG100} both $M_*(M_\rmn{dyn}, \text{TNG50})$ and $M_*(M_\rmn{dyn}, \text{TNG100-1})$ represent the average stellar mass at the dynamical mass considered. We follow Equations~\eqref{eq:rescaleTNG300_toTNG50} and \eqref{eq:rescaleTGN100_toTNG50} up to dynamical masses of $10^{12.2}~\rmn{M}_\odot$ for centrals, $10^{11.9}~\rmn{M}_\odot$ for satellites in hosts of $10^{12} - 10^{13}~\rmn{M}_\odot$, and $10^{11.6}~\rmn{M}_\odot$ for satellites in hosts of both $10^{13} - 10^{14}~\rmn{M}_\odot$ and $10^{14} - 10^{15.2}~\rmn{M}_\odot$. At larger dynamical mass, statistics in TNG50 become insufficient to continue the rescaling process in the same way. In order to avoid a sharp drop in stellar mass to the level of TNG100-1, we still include TNG50 galaxies at higher dynamical masses. However, since galaxies in this mass range are subject to sample variance, we only use the median stellar mass of all galaxies within a larger dynamical mass bin of $0.7~\rmn{dex}$ to a power of 0.5.

\section{Fitting the satellite stellar-to-halo mass relation as a function of host mass}
\label{sec:app_fits}

\begin{figure*}
    \centering
    \includegraphics[width=.8\textwidth]{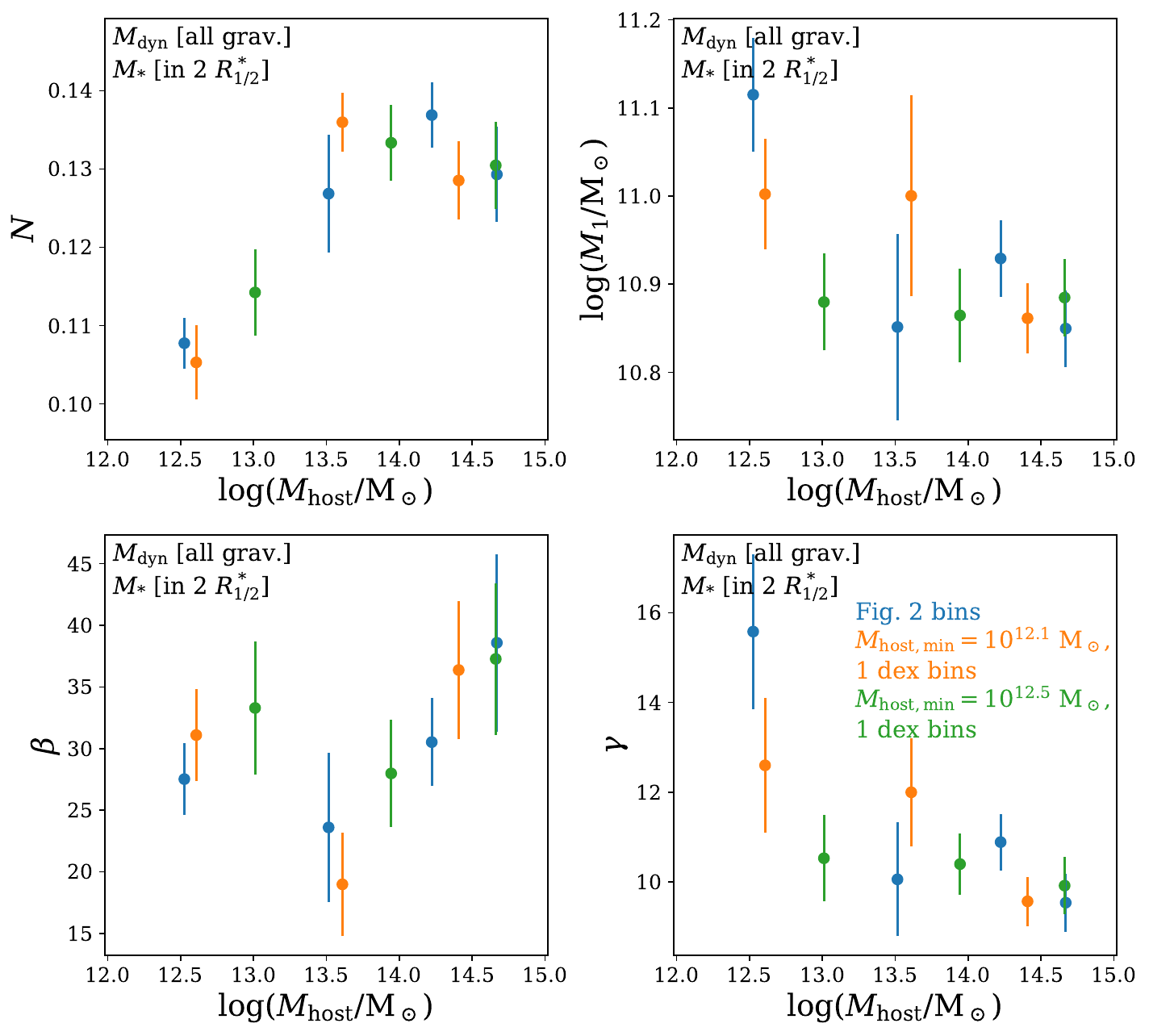}
    \caption{Fit parameters for the satellite SHMR as a function of median host mass. We use Equation~\eqref{eq:mosterfit} to fit the SHMR. This includes the normalisation $N$ (upper left panel), characteristic mass $M_1$ (upper right panel), and the slopes at the low- and high-mass ends $\beta$ and $\gamma$ (lower left and right panel, respectively). In order to illustrate their dependence on host mass, we show the fit parameters for satellites in different bins of host mass. Errorbars correspond to the respective fitting errors.}
    \label{fig:fit_vs_Mhost}
\end{figure*}

In Section~\ref{sec:tools}, we provide fitting formulas for the SHMR of centrals and satellites considering various bins of host mass. The dependence of the four fitting parameters -- normalisation $N$, characteristic mass $M_1$, and the slopes at the low- and high-mass ends $\beta$ and $\gamma$ -- on host mass is illustrated in Figure~\ref{fig:fit_vs_Mhost}. Masses used in the SHMR are measured in our fiducial aperture choice: all gravitationally bound particles for $M_\rmn{dyn}$, and stellar mass with twice the stellar half-mass radius for $M_*$.


\bsp	
\label{lastpage}
\end{document}